\documentclass[twocolumn,pre,aps,showpacs,showkeys,amsmath]{revtex4}
\usepackage{graphicx}
\usepackage{adjustbox}
\usepackage{multirow}

\begin{document}

\title{Population and Individual Firing Behaviors in Sparsely Synchronized Rhythms in The Hippocampal Dentate Gyrus}
\author{Sang-Yoon Kim}
\email{sykim@icn.re.kr}
\author{Woochang Lim}
\email{wclim@icn.re.kr}
\affiliation{Institute for Computational Neuroscience and Department of Science Education, Daegu National University of Education, Daegu 42411, Korea}

\begin{abstract}
We investigate population and individual firing behaviors in sparsely synchronized rhythms (SSRs) in a spiking neural network of the hippocampal dentate gyrus (DG). The main encoding granule cells (GCs) are grouped into lamellar clusters. In each GC cluster, there is one inhibitory (I) basket cell (BC) along with excitatory (E) GCs, and they form the E-I loop. Winner-take-all competition, leading to sparse activation of the GCs, occurs in each GC cluster. Such sparsity has been thought to enhance pattern separation performed in the DG. During the winner-take-all competition, SSRs are found to appear in each population of the GCs and the BCs through interaction of excitation of the GCs with inhibition of the BCs. Sparsely synchronized spiking stripes appear successively with the population frequency $f_p~ (= 13$ Hz) in the raster plots of spikes. We also note that excitatory hilar mossy cells (MCs) control the firing activity of the GC-BC loop by providing excitation to both the GCs and the BCs. SSR also appears in the population of MCs via interaction with the GCs (i.e., GC-MC loop). Population behaviors in the SSRs are quantitatively characterized in terms of the synchronization measures. In addition, we investigate individual firing activity of GCs, BCs, and MCs in the SSRs. Individual GCs exhibit random spike skipping, leading to a multi-peaked inter-spike-interval histogram, which is well characterized in terms of the random phase-locking degree. In this case, population-averaged mean-firing-rate (MFR) $\langle f_i^{(\rm GC)} \rangle$ is less than the population frequency $f_p$. On the other hand, both BCs and MCs show ``intrastripe'' burstings within stripes, together with ``interstripe'' random spike skipping. Thus, the population-averaged MFR $\langle f_i^{(X)} \rangle$ ($X=$ MC and BC) is larger than $f_p$, in contrast to the case of the GCs.  MC loss may occur during epileptogenesis. With decreasing the fraction of the MCs, changes in the population and individual firings in the SSRs are also studied. Finally, quantitative association between the population/individual firing behaviors in the SSRs and the winner-take-all competition is discussed.
\end{abstract}


\maketitle

\section{Introduction}
\label{sec:INT}
The hippocampus, composed of the dentate gyrus (DG) and the areas CA3 and CA1, plays important roles in memory formation, storage, and retrieval
\cite{Gluck,Squire}. The DG is the first subregion of the hippocampus, and its primary granule cells (GCs) receive excitatory inputs from the entorhinal cortex (EC)
through the perforant paths (PPs). As a pre-processor for the CA3, the GCs perform pattern separation on the input patterns from the EC by sparsifying and orthogonalizing them (i.e., transforming a set of input patterns into sparser and orthogonalized patterns), and project the pattern-separated outputs to the pyramidal cells in the CA3 via the mossy fibers (MFs) \cite{Marr,Will,Mc,Rolls1,Rolls2a,Rolls2b,Treves1,Treves2,Treves3,Oreilly,Schmidt,Rolls3,Knier,Myers1,Myers2,Scharfman,Yim,Chavlis,PS1,PS2,PS3,PS4,PS5,PS6,PS7}.
The sparse, but relatively strong MFs are known to play a role of ``teaching inputs'' which tend to trigger synaptic plasticity between the pyramidal cells in the CA3 and also between the pyramidal cells and the EC cells \cite{Treves3,Oreilly,Rolls3,Myers2,Scharfman}. Then, a new pattern may be stored in modified synapses. In this way, pattern separation in the DG facilitates pattern storage in the CA3.

The whole GCs are grouped into the lamellar clusters \cite{Cluster1,Cluster2,Cluster3,Cluster4}. In each GC cluster, there is one inhibitory (I) basket cell (BC)
along with excitatory (E) GCs, and they form a dynamical E-I loop. During the process of pattern separation, the GCs make sparse firing activity via the winner-take-all competition \cite{Rolls1,Treves3,Rolls3,Myers1,WTA1,WTA2,WTA3,WTA4,WTA5,WTA6,WTA7,WTA8,WTA9,WTA10}. Only strongly active GCs survive under the feedback inhibition of the BC (i.e., they become winners), while weakly active GCs become silent in response to the feedback inhibition from the BC.
The sparsity has been thought to enhance the pattern separation \cite{Treves3,Oreilly,Schmidt,Rolls3,Myers1,Myers2,Scharfman,Yim,Chavlis}.

Here, we are concerned about population rhythms in the DG. For example, gamma rhythms were observed to appear for communication between the DG and the EC, and between the DG and the CA3 \cite{DGR1,DGR2}. Also, decrease in the amplitude of theta rhythm and increase in the amplitude of beta rhythm were observed in the DG while performing different associative tasks via presentation of meaningful cues \cite{DGR3}. In this paper, we consider a spiking neural network of the hippocampal DG. During the winner-take-all competition, sparsely synchronized rhythms (SSRs) are found to emerge in each population of the GCs and the BCs via interaction of excitation of the GCs and inhibition of the BCs. Similar sparsely-synchronized population rhythms were observed in the hippocampus, the neocortex, the cerebellum, and the olfactory system \cite{FSS1,FSS2,FSS3,FSS4}.

We investigate the population and individual behaviors in the SSRs. Sparsely synchronized stripes (composed of spikes and indicating population synchronization) appear successively in the raster plot of spikes, and the corresponding instantaneous spike rates (IPSRs) exhibit oscillatory behaviors with the population frequency $f_p~(= 13$ Hz). In addition to the excitatory GCs, there exist another type of excitatory hilar mossy cells (MCs), in contrast to the case of the CA3 and the CA1 with only one type of excitatory pyramidal cells. The MCs control the firing activity of the GC-BC loop by providing excitation to both the GCs and the BCs. SSR is also found to appear in the population of MCs via interaction with the GCs (i.e., GC-MC loop). Thus, in the whole DG network, SSRs appear in the populations of the GCs, the MCs, and the BCs, together with occurrence of the winner-take-all competition, leading to sparse activation of the GCs (enhancing the pattern separation).

Population behaviors in the SSRs are quantitatively characterized by employing diverse synchronization measures introduced in our prior works.
The overall synchronization degree for the SSR may be well measured in terms of a thermodynamic amplitude measure, given by the time-averaged amplitude of the
macroscopic IPSR \cite{AM}. In addition, we use the statistical-mechanical spiking measure, given by the product of the occupation degree (representing the
spike density in each stripe) and the pacing degree between the spikes \cite{SM}, and make intensive characterization of the SSRs.

We also study the individual firing behaviors of the GCs, the MCs, and the BCs in the SSRs. Active GCs exhibit intermittent spikings, phase-locked to the
IPSR at random multiples of the global period of the IPSR. As a result, the inter-spike-interval (ISI) histogram consists of ``interstripe'' skipping peaks.
Similar skipping phenomena of spikings (characterized with multi-peaked ISI histograms) were also observed in the case of fast sparse synchronization
occurring in the systems consisting of the two excitatory and inhibitory populations or in the single inhibitory population \cite{W_Review,Sparse1,Sparse2,Sparse3,FSS}.
Due to random spike skipping, population-averaged mean-firing-rate (MFR) $\langle f_i^{(\rm GC)} \rangle~(=$ 2 Hz) becomes less than the population frequency $f_p$.
We also introduce a new random phase-locking degree and characterize the random spike skipping.
In contrast to the GCs, both MCs and BCs exhibit “intrastripe” burstings within stripes, along with ``interstripe'' random spike skipping.
Hence, the ISI histogram becomes composed of the dominant intrastripe bursting peak and  the interstripe skipping peaks.
Due to the dominant intrastripe bursting peak, the population-averaged MFR $\langle f_i^{(X)} \rangle$ ($X=$ MC and BC) is larger than $f_p$, in contrast to the case of the GCs.

Finally, we note that death of hilar MCs may occur during epileptogenesis \cite{BN1,BN2}. We decrease the fraction of the MCs, and study how the population and individual firings in the SSRs change. Quantitative correlation between the population/individual firing behaviors in the SSRs and the winner-take-all competition is also studied.

This paper is organized as follows. In Sec.~\ref{sec:DGN}, we describe a spiking neural network of the hippocampal DG. Then, in the main Sec.~\ref{sec:SSR}, we investigate population and individual behaviors in the SSRs of the GCs, the MCs, and the BCs. Finally, we give summary and discussion in Sec.~\ref{sec:SUM}.

\section{Spiking Neural Network of The Dentate Gyrus}
\label{sec:DGN}
In this section, we describe our spiking neural network of the DG. By following our prior approach for the cerebellar spiking neural network \cite{Kim1,Kim2},
we first developed our DG spiking neural network in the work for the winner-take-all competition \cite{WTA}, based on the anatomical and the physiological properties described in \cite{Myers1,Chavlis}. The system parameters for the structure, the single neuron models, and the synaptic currents in the present work are the same as those in the work for the winner-take-all competition; for details on our DG network, refer to Sec. 2 in \cite{WTA}.

\begin{figure}
\includegraphics[width=\columnwidth]{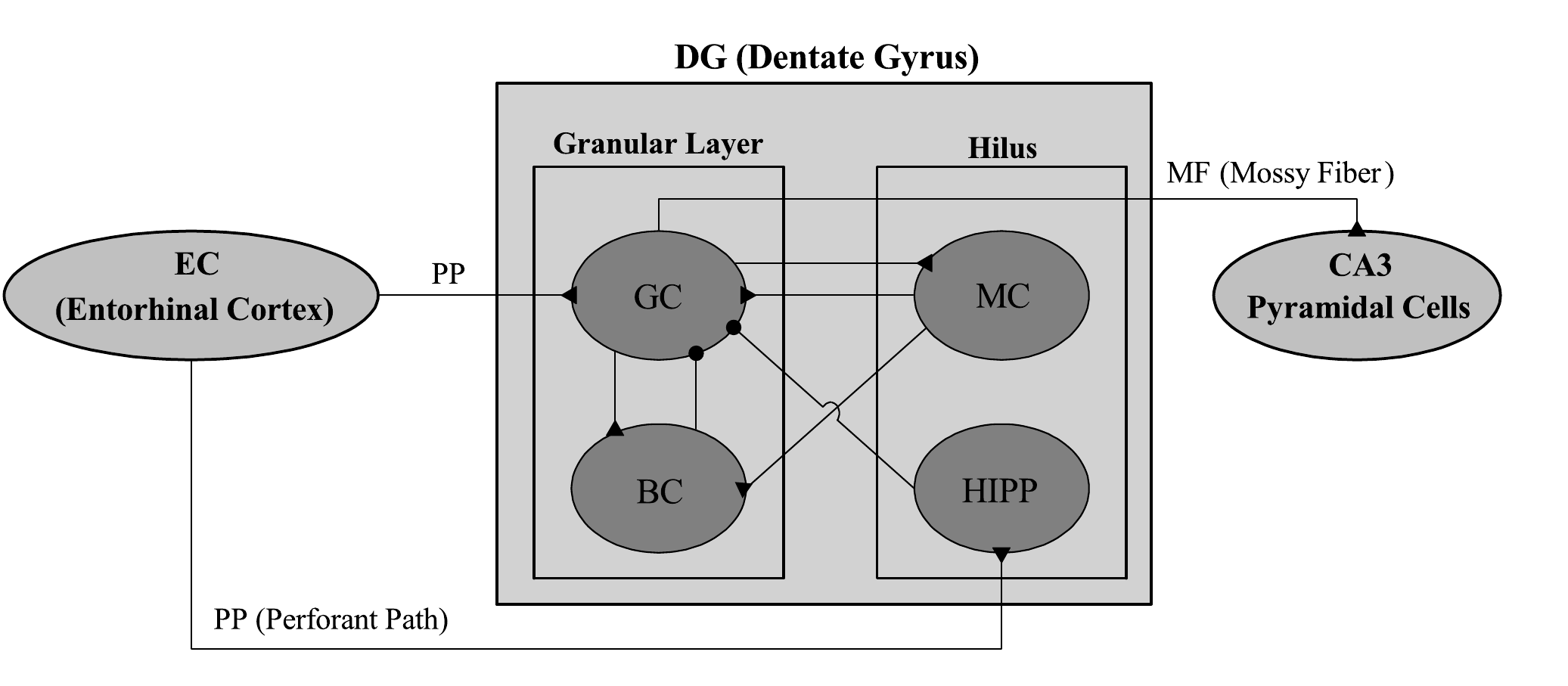}
\caption{Box diagram for the hippocampal dentate gyrus (DG) network. Lines with triangles and circles denote excitatory and inhibitory synapses, respectively. In the
DG, there are the granular layer [consisting of GC (granule cell) and BC (basket cell)] and the hilus [composed of MC (mossy cell) and HIPP (hilar perforant path-associated) cell]. The DG receives excitatory input from the EC (entorhinal cortex) via PPs (perforant paths) and provides its output to the CA3 via
MFs (mossy fibers).
}
\label{fig:DGN}
\end{figure}

\subsection{Framework of The Spiking Neural Network of The Dentate Gyrus}
\label{subsec:SNN}
Figure \ref{fig:DGN} shows the box diagram for our DG network. The granular layer (composed of the excitatory GCs and the inhibitory BCs) and the hilus [consisting of the excitatory MCs and the inhibitory HIPP (hilar perforant path-associated) cells] constitute the DG. Thus, there exist two types of excitatory cells, GCs and MCs, in contrast to the case of the CA3 and CA1 with only one kind of excitatory pyramidal cells. This DG receives the input from the external EC via the PPs and projects its output to the CA3 via the MFs.

As in \cite{Myers1,Chavlis}, based on the anatomical data, we choose the numbers of the constituent cells (GCs, BCs, MCs, and HIPP cells) in the DG and the EC cells and the connection probabilities between them. In \cite{WTA}, we developed a scaled-down spiking neural network where the total number of excitatory GCs  ($N_{\rm GC}$) was 2,000, corresponding to $\frac {1}{500}$ of the $10^6$ GCs found in rats \cite{ANA1}. These GCs were grouped into the $N_c~(=100)$ lamellar clusters \cite{Cluster1,Cluster2,Cluster3,Cluster4}. In each GC cluster, there were $n_{\rm {GC}}^{(c)}~(=20)$ GCs and one inhibitory BC. Hence, the number of BCs
($N_{\rm BC}$) in the whole DG network became 100, corresponding to $\frac {1} {20}$ of $N_{\rm GC}$. Thus, in each GC cluster, a dynamical GC-BC loop was formed, and the BC (receiving the excitation from all the GCs) provided the feedback inhibition to all the GCs.

The EC layer II is the external source providing the excitatory inputs to the GCs and the HIPP cells via the PPs. The estimated number of the EC layer II cells ($N_{\rm EC}$) is about 200,000 in rats, corresponding to 20 EC cells per 100 GCs  \cite{ANA3}. Thus, we chose $N_{\rm EC}=400$ in our DG network.
Also, the activation degree $D_a$ of the EC cells was chosen as 10$\%$ \cite{ANA4}. Thus, we randomly chose 40 active ones among the 400 EC (layer II) cells.
Each active EC cell was modeled in terms of the Poisson spike train with frequency of 40 Hz \cite{ANA5}.
The random-connection probability $p^{\rm (GC,EC)}$ ($p^{\rm (HIPP,EC)}$) from the pre-synaptic EC cells to a post-synaptic GC (HIPP cell) was 20 $\%$. Thus, each GC or HIPP cell was randomly connected with the average number of 80 EC cells.

Next, we consider the hilus, composed of the excitatory MCs and the inhibitory HIPP cells \cite{Myers1,Chavlis,Yim,Hilus1,Hilus2,Hilus3,Hilus4,Hilus5,Hilus6,Hilus7}.
In rats, the number of MCs ($N_{\rm MC}$) is known to change from 30,000 to 50,000, corresponding to 3-5 MCs per 100 GCs \cite{ANA2}. In our DG network, we chose $N_{\rm MC}~=80$. Also, the estimated number of HIPP cells ($N_{\rm HIPP}$) is about 12,000, which corresponds to 2 HIPP cells per 100 GCs. Hence, we chose $N_{\rm HIPP}=40$ in our DG network.

In our DG network, the hilar MCs and the GCs were mutually connected with 20 $\%$ random-connection probabilities $p^{\rm (MC,GC)}$ ($\rm GC \rightarrow MC$) and $p^{\rm (GC,MC)}$ ($\rm MC \rightarrow GC$). In this way, the GCs and the MCs formed a dynamical E-E loop.  All the MCs also provided the excitation to the BC in each GC cluster. Hence, the BC in the GC cluster received excitatory inputs from all the GCs in the same GC cluster and from all the MCs. In this way, the MCs control the firing activity in the GC-BC loop by providing excitation to both the GCs and the BCs.

We also note that each GC in the GC cluster received inhibition from the randomly-connected HIPP cells with the connection probability $p^{\rm (GC,HIPP)}~=~20 ~\%$.
Hence, the firing activity of the GCs may be determined through competition between the excitatory inputs from the EC cells and from the MCs and the inhibitory inputs from the HIPP cells.

With the above information on the numbers of the relevant cells and the connection probabilities between them, we developed a one-dimensional ring network for the DG \cite{WTA}. As in the famous small-world ring network \cite{SWN1,SWN2}, our ring network has advantage for computational and analytical efficiency, and its visual representation may also be easily made. For the schematic diagrams of the ring networks for the EC, the granular layer and the hilus, refer to Figs.~1(b1)-1(b3)
in \cite{WTA}, respectively.

\subsection{Elements and Synaptic Currents in The DG Spiking Neural Network}
\label{subsec:LIF-SC}
As elements of our DG spiking neural network, we chose leaky integrate-and-fire (LIF) neuron models with additional afterhyperpolarization (AHP) currents which  determines refractory periods, like our prior study of cerebellar network \cite{Kim1,Kim2,WTA}. This LIF neuron model is one of the simplest spiking neuron models \cite{LIF}. Due to its simplicity, it may be easily analyzed and simulated.

Evolutions of dynamical states of individual cells in the $X$ population are governed by the following equations:
\begin{equation}
C_{X} \frac{dv_{i}^{(X)}}{dt} = -I_{L,i}^{(X)} - I_{AHP,i}^{(X)} + I_{ext}^{(X)} - I_{syn,i}^{(X)}, \;\;\; i=1, \cdots, N_{X}.
\label{eq:GE}
\end{equation}
Here, $N_X$ is the total number of cells in the $X$ population, $X=$ GC and BC in the granular layer and $X=$ MC and HIPP in the hilus.
In Eq.~(\ref{eq:GE}), $C_{X}$ (pF) represents the membrane capacitance of the cells in the $X$ population, and the state of the $i$th cell in the $X$ population at a time $t$ (msec) is characterized by its membrane potential $v_i^{(X)}$ (mV). The time-evolution of $v_i^{(X)}(t)$ is governed by 4 types of currents (pA) into the
$i$th cell in the $X$ population; the leakage current $I_{L,i}^{(X)}$, the AHP current $I_{AHP,i}^{(X)}$, the external constant current $I_{ext}^{(X)}$ (independent of $i$), and the synaptic current $I_{syn,i}^{(X)}$. Here, a subthreshold case of $I_{ext}^{(X)}=0$ was considered for all $X$ \cite{Chavlis}.

The 1st type of leakage current $I_{L,i}^{(X)}$ for the $i$th cell in the $X$ population is given by:
\begin{equation}
I_{L,i}^{(X)}(t) = g_{L}^{(X)} (v_{i}^{(X)}(t) - V_{L}^{(X)}).
\label{eq:Leakage}
\end{equation}
Here, $g_L^{(X)}$ and $V_L^{(X)}$ are conductance (nS) and reversal potential for the leakage current, respectively.
The $i$th cell fires a spike when its membrane potential $v_i^{(X)}$ reaches a threshold $v_{th}^{(X)}$ at a time $t_{f,i}^{(X)}$.
Then, the 2nd type of AHP current $I_{AHP,i}^{(X)}$ follows after spiking (i.e., $t \geq t_{f,i}^{(X)}$), :
\begin{equation}
I_{AHP,i}^{(X)}(t) = g_{AHP}^{(X)}(t) ~(v_{i}^{(X)}(t) - V_{AHP}^{(X)})~~~{\rm ~for~} \; t \ge t_{f,i}^{(X)}.
\label{eq:AHP1}
\end{equation}
Here, $V_{AHP}^{(X)}$ is the reversal potential for the AHP current, and the conductance $g_{AHP}^{(X)}(t)$ is given by an exponential-decay
function:
\begin{equation}
g_{AHP}^{(X)}(t) = \bar{g}_{AHP}^{(X)}~  e^{-(t-t_{f,i}^{(X)})/\tau_{AHP}^{(X)}}.
\label{eq:AHP2}
\end{equation}
Here, $\bar{g}_{AHP}^{(X)}$ and $\tau_{AHP}^{(X)}$ are the maximum conductance and the decay time constant for the AHP current.
As $\tau_{AHP}^{(X)}$ is increased, the refractory period becomes longer.

For the parameter values of the capacitance $C_X$, the leakage current $I_L^{(X)}$, and the AHP current, refer to Table 1 in \cite{WTA};
these parameter values are based on physiological properties of the GC, BC, MC, and HIPP cell \cite{Chavlis,Hilus3}.

We next consider the synaptic current $I_{syn,i}^{(X)}$ into the $i$th cell in the $X$ population, composed of the following 3 types of synaptic currents:
\begin{equation}
I_{syn,i}^{(X)} = I_{{\rm AMPA},i}^{(X,Y)} + I_{{\rm NMDA},i}^{(X,Y)} + I_{{\rm GABA},i}^{(X,Z)}.
\label{eq:ISyn1}
\end{equation}
Here, $I_{{\rm AMPA},i}^{(X,Y)}$ and $I_{{\rm NMDA},i}^{(X,Y)}$ are the excitatory AMPA ($\alpha$-amino-3-hydroxy-5-methyl-4-isoxazolepropionic acid) receptor-mediated and NMDA ($N$-methyl-$D$-aspartate) receptor-mediated currents from the pre-synaptic source $Y$ population to the post-synaptic $i$th neuron in the target $X$ population. On the other hand, $I_{{\rm GABA},i}^{(X,Z)}$ is the inhibitory $\rm GABA_A$ ($\gamma$-aminobutyric acid type A) receptor-mediated current
from the pre-synaptic source $Z$ population to the post-synaptic $i$th neuron in the target $X$ population.

Like the case of the AHP current, the $R$ (= AMPA, NMDA, or GABA) receptor-mediated synaptic current $I_{R,i}^{(T,S)}$ from the pre-synaptic source $S$ population to the $i$th post-synaptic cell in the target $T$ population is given by:
\begin{equation}
I_{R,i}^{(T,S)}(t) = g_{R,i}^{(T,S)}(t)~(v_{i}^{(T)}(t) - V_{R}^{(S)}).
\label{eq:ISyn2}
\end{equation}
Here, $g_{(R,i)}^{(T,S)}(t)$ and $V_R^{(S)}$ are synaptic conductance and synaptic reversal potential
(determined by the type of the pre-synaptic source $S$ population), respectively.

We get the synaptic conductance $g_{R,i}^{(T,S)}(t)$ from:
\begin{equation}
g_{R,i}^{(T,S)}(t) = K_{R}^{(T,S)} \sum_{j=1}^{N_S} w_{ij}^{(T,S)} ~ s_{j}^{(T,S)}(t).
\label{eq:ISyn3}
\end{equation}
Here, $K_{R}^{(T,S)}$ is the synaptic strength per synapse for the $R$-mediated synaptic current
from the $j$th pre-synaptic neuron in the source $S$ population to the $i$th post-synaptic cell in the target $T$ population.
The inter-population synaptic connection from the source $S$ population (with $N_s$ cells) to the target $T$ population is given by the connection weight matrix
$W^{(T,S)}$ ($=\{ w_{ij}^{(T,S)} \}$) where $w_{ij}^{(T,S)}=1$ if the $j$th cell in the source $S$ population is pre-synaptic to the $i$th cell
in the target $T$ population; otherwise $w_{ij}^{(T,S)}=0$.

The post-synaptic ion channels are opened because of binding of neurotransmitters (emitted from the source $S$ population) to receptors in the target
$T$ population. The fraction of open ion channels at time $t$ is represented by $s^{(T,S)}$. The time course of $s_j^{(T,S)}(t)$ of the $j$th cell
in the source $S$ population is given by a sum of double exponential functions $E_{R}^{(T,S)} (t - t_{f}^{(j)}-\tau_{R,l}^{(T,S)})$:
\begin{equation}
s_{j}^{(T,S)}(t) = \sum_{f=1}^{F_{j}^{(s)}} E_{R}^{(T,S)} (t - t_{f}^{(j)}-\tau_{R,l}^{(T,S)}).
\label{eq:ISyn4}
\end{equation}
Here, $t_f^{(j)}$ and $F_j^{(s)}$ are the $f$th spike time and the total number of spikes of the $j$th cell in the source $S$ population, respectively, and
$\tau_{R,l}^{(T,S)}$ is the synaptic latency time constant for $R$-mediated synaptic current.
The exponential-decay function $E_{R}^{(T,S)} (t)$ (corresponding to contribution of a pre-synaptic spike occurring at $t=0$ in the absence of synaptic latency)
is given by:
\begin{equation}
E_{R}^{(T,S)}(t) = \frac{1}{\tau_{R,d}^{(T,S)}-\tau_{R,r}^{(T,S)}} \left( e^{-t/\tau_{R,d}^{(T,S)}} - e^{-t/\tau_{R,r}^{(T,S)}} \right) \cdot \Theta(t). \label{eq:ISyn5}
\end{equation}
Here, $\Theta(t)$ is the Heaviside step function: $\Theta(t)=1$ for $t \geq 0$ and 0 for $t <0$, and $\tau_{R,r}^{(T,S)}$ and $\tau_{R,d}^{(T,S)}$ are synaptic rising and decay time constants of the $R$-mediated synaptic current, respectively.

For the  parameter values for the synaptic strength per synapse $K_{R}^{(T,S)}$, the synaptic rising time constant $\tau_{R,r}^{(T,S)}$, synaptic decay time constant $\tau_{R,d}^{(T,S)}$, synaptic latency time constant $\tau_{R,l}^{(T,S)}$, and the synaptic reversal potential $V_{R}^{(S)}$, refer to Tables 2 and 3 in \cite{WTA}.
These parameter values are also based on the physiological properties of the relevant neurons \cite{Chavlis,SynParm1,SynParm2,SynParm3,SynParm4,SynParm5,SynParm6,SynParm7,SynParm8}.

\section{Population and Individual Firing Behaviors in Sparsely Synchronized Rhythms}
\label{sec:SSR}
In our DG network shown in Fig.~\ref{fig:DGN}, the main encoding GCs are found to exhibit SSR during their winner-take-all competition
(leading to sparse activation of the GCs). Also, the MCs and the BCs show SSRs via interaction with the GCs through the GC-MC loop and the GC-BC loop.
Population and individual behaviors of these SSRs are investigated in terms of diverse measures for population synchronization and random phase-locking degree
for characterization of the ISIs. Quantitative correlations between these behaviors and the winner-take-all competition are shown to exist.

\subsection{Population and Individual Behaviors in the Sparsely Synchronized Rhythm of the GCs}
\label{subsec:GCR}
Figure \ref{fig:DGN} shows the external input from the EC. There are direct excitatory input from the EC cells and indirect disynaptic inhibitory EC input, mediated by the HIPP cells. Thus, the EC cells and the HIPP cells become the excitatory and the inhibitory input sources to the GCs, respectively.

Among the 400 EC cells, randomly-chosen 40 active cells make spikings (i.e., activation degree $D_a=10$ $\%$). Each active EC cell
is modeled in terms of the Poisson spike train with frequency of 40 Hz. After a break stage ($t=0-300$ msec), Poisson spike train of each active EC cell
follows during the stimulus stage ($t=300- 30,300$ msec; the stimulus period $T_s$ is $3 \cdot 10^4$ msec).

We note that each HIPP cell is randomly connected to the average number of 80 EC cells with the connection probability $p^{\rm (HIPP,EC)}$ = 20$\%$, among which the average number of active EC cells is 8. Among the 40 HIPP cells, 37 HIPP cells are found to be active, while the remaining 3 HIPP cells (without receiving excitatory input from the active EC cells) are silent; the activation degree of the HIPP cells is 92.5$\%$. Also, the spikings of the active HIPP cells begin from $t \simeq 320$ msec (i.e. about 20 msec delay for the firing of the HIPP cells with respect to the firing onset ($t=300$ msec) of the active EC cells).

\begin{figure}
\includegraphics[width=0.8\columnwidth]{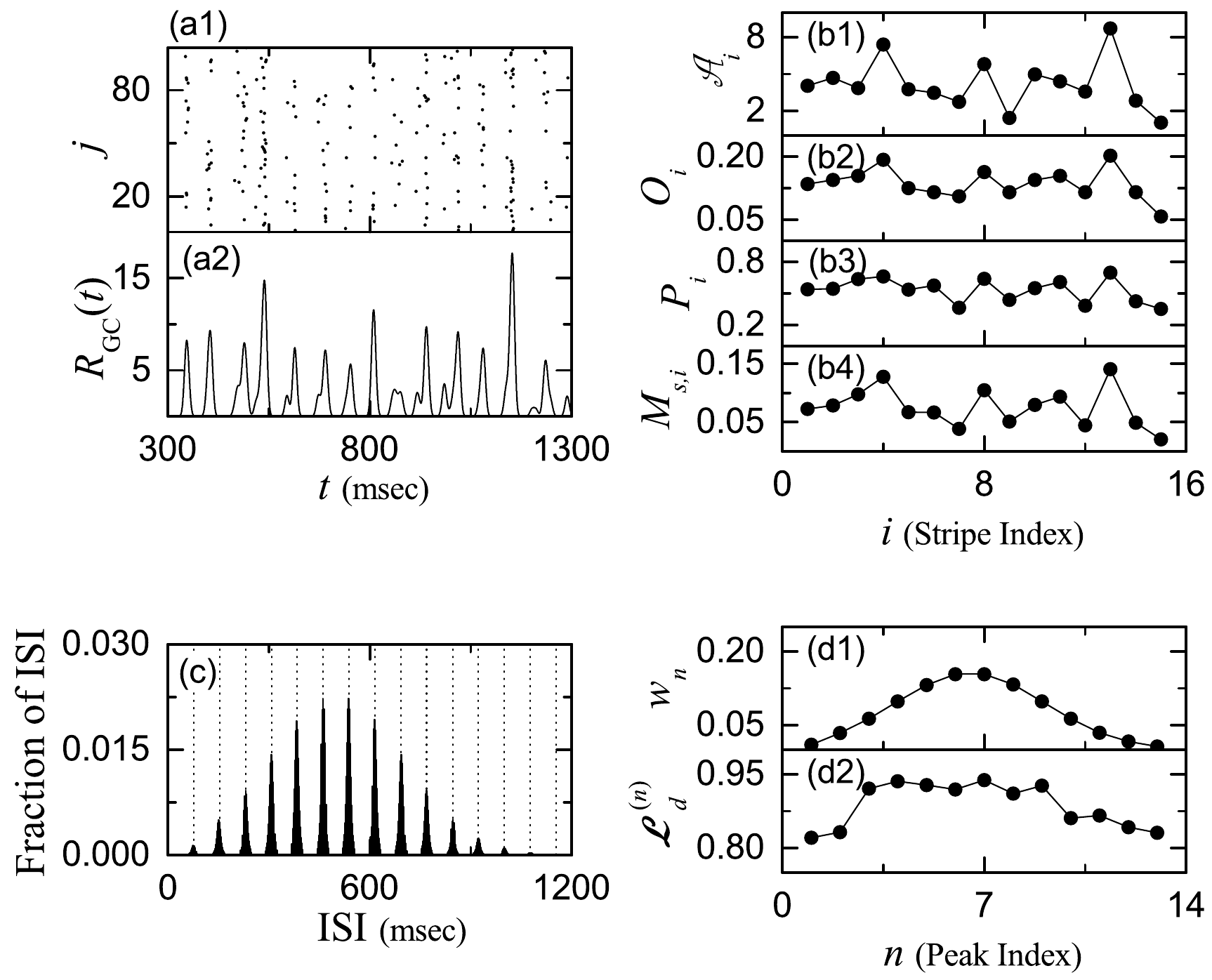}
\caption{Emergence of SSR of the GCs. (a1) Raster plot of spikes of 104 active GCs. (a2) IPSR $R_{\rm GC}(t)$ of active GCs. Band width for $R_{\rm GC}(t)$: $h=20$ msec. Plots of (b1) amplitude ${\cal A}_i$ of the IPSR $R_{\rm GC}(t)$, (b2) occupation degree $O_i$, (b3) pacing degree $P_i$, and (b4) statistical-mechanical spiking measure $M_{s,i}$ vs. $i$ (spiking stripe index). (c) Population-averaged ISI histogram; bin size = 2 msec. Vertical dotted lines in (c) represent the integer multiples of the global period $T_G^{\rm (GC)}$ (=76.9 msec) of  $R_{\rm GC}(t)$. Plots of (d1) normalized weight $w_n$ and (d2) random phase-locking degree ${\cal L}_d^{(n)}$ for the $n$th peak of the ISI histogram versus $n$ (skipping peak index).
}
\label{fig:GC}
\end{figure}

As a pre-processor for the CA3, the GCs in the DG perform the pattern separation, facilitating the pattern storage in the CA3.
The GCs make sparse firing activity through competitive learning, which has been thought to improve the pattern separation.
The activation degree of the GCs was found to be $D_a=5.2\%$ (i.e., the total number of active GCs is 104) \cite{WTA}.
Also, the active GCs begin to make sparse firings from $t \simeq 340$ msec [i.e. about 40 msec delay for the firing of the GCs with respect to the firing onset ($t=300$ msec) of the active EC cells]. Dynamical origin for winner-take-all competition, leading to the sparse activation of the GCs, has been studied in our prior work \cite{WTA}. Winner-take-all competition has been found to occur via competition between the firing activity of the GCs and the feedback inhibition of the BC in each GC cluster. In this case, the hilar MCs has also been found to enhance the winner-take-all competition by providing excitation to both the GCs and the BC.

During the winner-take-all competition, SSR is found to appear in the population of the GCs via interaction of excitation of the GCs with inhibition of the BCs.
Population firing activity of the active GCs may be well visualized in the raster plot of spikes which is a collection of spike trains of individual active GCs.
Figure \ref{fig:GC}(a1) shows the raster plot of spikes for the active GCs; for convenience, only a part from $t=300$ to 1,300 msec is shown in the raster plot of spikes. We note that sparsely synchronized stripes (composed of spikes and indicating population sparse synchronization) appear successively.

As a population quantity showing collective behaviors, we use an instantaneous population spike rate (IPSR) which may be obtained from the raster plots of spikes \cite{SM,W_Review,Sparse1,Sparse2,Sparse3,FSS}. To get a smooth IPSR, we employ the kernel density estimation (kernel smoother) \cite{Kernel}. Each spike in the raster plot is convoluted (or blurred) with a kernel function $K_h(t)$ to get a smooth estimate of IPSR $R_{\rm GC}(t)$:
\begin{equation}
R_{\rm{GC}}(t) = \frac{1}{N_a} \sum_{i=1}^{N_a} \sum_{s=1}^{n_i} K_h (t-t_{s}^{(i)}),
\label{eq:IPSR}
\end{equation}
where $N_a$ is the number of the active GCs, $t_{s}^{(i)}$ is the $s$th spiking time of the $i$th active GC, $n_i$ is the total number of spikes for the $i$th active GC, and we use a Gaussian kernel function of band width $h$:
\begin{equation}
K_h (t) = \frac{1}{\sqrt{2\pi}h} e^{-t^2 / 2h^2}, ~~~~ -\infty < t < \infty.
\label{eq:Gaussian}
\end{equation}
Throughout the paper, the band width $h$ of $K_h(t)$ is 20 msec. The IPSR $R_{\rm {GC}}(t)$ of the active GCs is shown in Fig.~\ref{fig:GC}(a2), and we note that
$R_{\rm {GC}}(t)$ exhibits synchronous oscillation with the population frequency $f_p^{\rm (GC)}~(= 13$ Hz); in the desynchronized case, the IPSR becomes stationary without oscillation.

In the above way, SSR with the population frequency $f_p^{\rm (GC)}~(= 13$ Hz) [i.e., the global period $T_G^{\rm (GC)}$ (corresponding to the average period between the neighboring spiking stripes) is 76.9 msec] emerges in the population of active GCs. This is similar to the previously-studied case where fast sparse synchronization occurs via E-I balance in the feedback E-I loop \cite{W_Review,Sparse1,Sparse2,Sparse3}.

We now characterize population firing behavior in the SSR of the GCs by employing the thermodynamic amplitude measure and the statistical-mechanical spiking measure
\cite{AM,SM}. The thermodynamic amplitude measure ${\cal M}_a$ is given by the time-averaged amplitude of the macroscopic IPSR $R_{\rm {GC}}(t)$ \cite{AM}:
\begin{equation}
  {\cal M}_a = {\overline {{\cal A}_i}};~ {\cal A}_i = \frac { R_{\rm GC, max}^{(i)}(t) - R_{\rm GC, min}^{(i)}(t) } {2},
\label{eq:Ma}
\end{equation}
where the overline represents time average, and $R_{\rm GC,max}^{(i)}(t)$ and $R_{\rm GC,min}^{(i)}(t)$ are the maximum and the minimum of $R_{\rm GC}(t)$ in its $i$th global cycle (corresponding to the $i$th spiking stripe). As ${\cal M}_a$ increases (i.e., the time-averaged amplitude of $R_{\rm GC}(t)$ is increased), the synchronization degree of the SSR becomes higher. Figure \ref{fig:GC}(b1) shows the plot of the amplitude ${\cal A}_i$ versus the spiking stripe index $i$. We follow the 389 stripes during the stimulus period $T_s$ ($ = 3 \cdot 10^4$ msec), and thus thermodynamic amplitude measure ${\cal M}_a$ (corresponding to the time-averaged amplitude $\overline { {\cal A}_i }$) is found to be 3.57.

Next, we characterize the population firing behaviors in terms of the statistical-mechanical spiking measure \cite{SM}.
For a synchronous case, spiking stripes appear successively in the raster plot of spikes. The spiking measure $M_{s,i}$ of the $i$th stripe is defined by the product of the occupation degree $O_i$ of spikes (denoting the spike density of the $i$th stripe) and the pacing degree $P_i$ of spikes (representing the degree of phase coherence between spikes in the $i$th stripe):
\begin{equation}
M_{s,i} = O_i \cdot P_i.
\label{eq:SMi}
\end{equation}
The occupation degree $O_i$ of spikes in the $i$th stripe is given by the fraction of spiking neurons:
\begin{equation}
   O_i = \frac {N_i^{(s)}} {N_a},
\label{eq:Oi}
\end{equation}
where $N_i^{(s)}$ is the number of spiking cells in the $i$th stripe, and $N_a$ is the total number of active cells (e.g., $N_a=104$ for the GCs).
In the case of sparse synchronization, $O_i<1$, in contrast to the case of full synchronization with $O_i=1$.

The pacing degree $P_i$ of spikes in the $i$th stripe can be determined in a statistical-mechanical way by considering their contributions to the macroscopic IPSR   $R_{\rm GC}(t)$. Central maxima of $R_{\rm GC}(t)$ between neighboring left and right minima of $R_{\rm GC}(t)$ coincide with centers of spiking stripes in the raster plot. A global cycle begins from a left minimum of $R_{\rm GC}(t)$, passes a maximum, and ends at a right minimum. An instantaneous global phase $\Phi(t)$ of $R_{GC}(t)$ was introduced via linear interpolation in the region forming a global cycle [for details, refer to Eqs.~(16) and (17) in \cite{SM}]. Then, the contribution of the $k$th microscopic spike in the $i$th stripe occurring at the time $t_k^{(s)}$ to $R_{\rm GC}(t)$ is given by $\cos \Phi_k$, where $\Phi_k$ is the global phase at the $k$th spiking time [i.e., $\Phi_k \equiv \Phi(t_k^{(s)})$]. A microscopic spike makes the most constructive (in-phase) contribution to $R_{\rm GC}(t)$ when the corresponding global phase $\Phi_k$ is $2 \pi n$ ($n=0,1,2, \dots$). In contrast, it makes the most destructive (anti-phase) contribution to $R_{\rm GC}(t)$ when $\Phi_k$ is $2 \pi (n-1/2)$. By averaging the contributions of all microscopic spikes in the $i$th stripe to $R_{\rm GC}(t)$, we get the pacing degree of spikes in the $i$th stripe [refer to Eq.~(18) in \cite{SM}]:
\begin{equation}
 P_i ={ \frac {1} {S_i}} \sum_{k=1}^{S_i} \cos \Phi_k,
\label{eq:Pi}
\end{equation}
where $S_i$ is the total number of microscopic spikes in the $i$th stripe.
Then, via averaging $M_{s,i}$ of Eq.~(\ref{eq:SMi}) over a sufficiently large number $N_s$ of stripes (e.g., $N_s=389$ for the GCs), we
obtain the statistical-mechanical spiking measure $M_s$ [refer to Eq.~(19) in \cite{SM}]:
\begin{equation}
M_s =  {\frac {1} {N_s}} \sum_{i=1}^{N_s} M_{s,i}.
\label{eq:SM}
\end{equation}

Figures~\ref{fig:GC}(b2)-\ref{fig:GC}(b4) show the plots of $O_i$, $P_i,$ and $M_{s,i}$, respectively, in the 15 spiking stripes in
Fig.~\ref{fig:GC}(a1). By following the 389 stripes during the stimulus period $T_s$ ($= 3 \cdot 10^4$ msec), we get the average occupation $\langle O_i \rangle~(=0.14)$, the average pacing degree $\langle P_i \rangle~(=0.45)$, and the statistical-mechanical spiking measure $M_s~(=0.063)$.
Since $\langle O_i \rangle$ is much less than 1, sparse synchronization occurs. In contrast, moderate pacing ($\langle P_i \rangle=0.45$) takes place between spikes in each stripe. Thus, the statistical-mechanical spiking measure $M_s~(=0.063)$, representing the overall synchronization degree, becomes so small,
mainly due to low occupation degree.

In addition to the population firing behavior, we also characterize individual spiking behaviors in the SSR. We obtain the ISI histogram for each active GC by collecting the ISIs during the stimulus period $T_s$ ($= 3 \cdot 10^4$ msec), and then get the population-averaged ISI histogram by averaging the individual ISI histograms for all the active GCs. Figure \ref{fig:GC}(c) shows the population-averaged ISI histogram. Each active GC exhibits intermittent spikings, phase-locked to $R_{\rm GC}(t)$ at random multiples of its global period $T_G^{\rm (GC)}~(=76.9$ msec). Due to the random spike skipping, distinct 13 multiple peaks appear at the integer multiples of $T_G^{(GC)}$ (denoted by the vertical dotted lines). This is in contrast to the case of full synchronization where only one dominant peak appears at the global period $T_G^{\rm (GC)}$; all cells fire regularly at each global cycle without skipping. Hereafter, these peaks will be called as the ``interstripe skipping'' peaks. The middle 6th- and 7th-order peaks are the highest ones, and hence spiking may occur most probably after 5- or 6-times spike skipping. This kind of structure in the ISI histogram is a little different from that in the case of fast sparse synchronization where the highest peak appears at the 1st-order peak, and then the heights of the higher-order peaks decrease successively \cite{W_Review,Sparse1,Sparse2,Sparse3,FSS}.

In the case of the active GCs, the average ISI ($\langle {\rm ISI} \rangle$) is 500.01 msec. Hence, the population-averaged MFR $\langle f_i^{\rm (GC)} \rangle$ (=$1/\langle {\rm ISI} \rangle$) is 2.0 Hz, which is much less than the population frequency $f_p^{\rm (GC)}~(=13$ Hz) of the SSR, in contrast to the case of full synchronization (with full occupation) where the population-averaged MFR is the same as the population frequency.

We introduce a new random phase-locking degree, denoting how well intermittent spikes make phase-locking to $R_{\rm GC}(t)$ at random multiples of its global period $T_G^{\rm (GC)}$, and characterize the degree of random spike skipping seen in the ISI histogram. By following the approach developed in the case of pacing degree
\cite{SM}, we introduce the random phase-locking degree to examine the regularity of individual firings (represented well in the sharpness of the interstripe skipping peaks).

We first locate the interstripe skipping peaks. The range of ISI in the $n$th-order peak is as follows:
\begin{widetext}
\begin{eqnarray}
&&(n-\frac {1}{2})~ T_G^{\rm (GC)} < {\rm ISI} < (n+ \frac {1}{2})~ T_G^{\rm (GC)}~~~~~{\rm for}~n \geq 2, \\
&&0 < {\rm ISI} < {\frac {3}{2}}~ T_G^{\rm (GC)}~~~~~{\rm for}~n=1.
\label{eq:Peak}
\end{eqnarray}
\end{widetext}
For each $n$th-order peak, we get the normalized weight $w_n$, given by:
\begin{equation}
w_n = \frac {N_{\rm ISI}^{(n)}} {N_{\rm ISI}^{(tot)}},
\end{equation}
where $N_{\rm ISI}^{(tot)}$ is the total number of ISIs obtained during the stimulus period ($T_s$ $= 3 \cdot 10^4$ msec) and $N_{\rm ISI}^{(n)}$ is the number of the ISIs in the $n$th-order peak. For the GCs, $N_{\rm ISI}^{(tot)} = 6,218$. Figure \ref{fig:GC}(d1) shows the plot of $w_n$ versus $n$ (peak index) for all the 13 peaks. The middle highest 6th and 7th-order peaks have $w_6=0.153$ and $w_7=0.154$.

We now consider the sequence of the ISIs, $\{ {\rm ISI}_i^{(n)},~ i=1,\dots,N_{\rm ISI}^{(n)} \}$, within the $n$th-order peak, and get the random
phase-locking degree ${\cal L}_d^{(n)}$ of the $n$th-order peak. Similar to the case of the pacing degree \cite{SM}, we provide a phase $\psi$ to each ${\rm ISI}_i^{(n)}$ via linear interpolation:
\begin{equation}
  \psi(\Delta {\rm ISI}_i^{(n)}) = \frac {\pi} {T_G^{\rm (GC)}}~\Delta {\rm ISI}_i^{(n)}~~~~~{\rm for}~n \geq 2,
\label{eq:phase1}
\end{equation}
where $\Delta {\rm ISI}_i^{(n)} = {\rm ISI}_i^{(n)} - n ~T_G^{\rm (GC) }$, leading to $- {\frac {T_G^{\rm (GC)}}{2}}  < \Delta {\rm ISI}_i^{(n)} < {\frac {T_G^{\rm (GC)}}{2}} $. However, for $n=1$, $\psi$ varies depending on whether the ISI lies in the left or the right part of the 1st-order peak:
\begin{widetext}
\begin{equation}
\psi(\Delta {\rm ISI}_i^{(1)}) = \left\{
\begin{array}{l}
\frac {\pi} {2~T_G^{\rm (GC)}}~\Delta {\rm ISI}_i^{(1)}~~~~~{\rm for}~ - {\frac {T_G^{\rm (GC)}}{2}} < {\rm ISI}_i^{(1)} < 0, \\
\frac {\pi} {T_G^{\rm (GC)}}~\Delta {\rm ISI}_i^{(1)}~~~~~{\rm for}~ 0 < {\rm ISI}_i^{(1)} < {\frac {T_G^{\rm (GC)}}{2}}.
\end{array}
\right.
\label{eq:phase2}
\end{equation}
\end{widetext}

Then, the contribution of the ${\rm ISI}_i^{(n)}$ to the locking degree ${\cal L}_d^{(n)}$ is given by $\cos ( \psi_i^{(n)})$;
$\psi_i^{(n)} = \psi (\Delta {\rm ISI}_i^{(n)})$. An ${\rm ISI}_i^{(n)}$ makes the most constructive contribution to ${\cal L}_d^{(n)}$ for $\psi_i^{(n)}=0$, while it makes no contribution to ${\cal L}_d^{(n)}$ for $\psi = {\frac {\pi} {2}}$ or $-{\frac {\pi} {2}}$.
By averaging the matching contributions of all the ISIs in the $n$th-order peak, we obtain:
\begin{equation}
  {\cal L}_d^{(n)} = { \frac {1}{N_{\rm ISI}^{(n)}} } \sum_i^{N_{\rm ISI}^{(n)}}  \cos ( \psi_i^{(n)}).
\label{eq:LDn}
\end{equation}

Finally, we get the (overall) random phase-locking degree ${\cal L}_d$ via weighted average of the random phase-locking degrees ${\cal L}_d^{(n)}$ of all the peaks:
\begin{equation}
  {\cal L}_d = {\frac {1} {N_p}} \sum_{n=1}^{N_p} w_n \cdot {\cal L}_d^{(n)}
             = {\frac {1} {N_{\rm ISI}^{\rm (tot)}} } \sum_{n=1}^{N_p} \sum_{i=1}^{N_{\rm ISI}^{\rm (tot)} } \cos ( \psi_i^{(n)}),
\label{eq:LD}
\end{equation}
where $N_p$ is the number of peaks in the ISI histogram.
Thus, ${\cal L}_d$ corresponds to the average of contributions of all the ISIs in the ISI histogram.
Figure \ref{fig:GC}(d2) shows the plot of ${\cal L}_d^{(n)}$ versus $n$ (peak index) for the 13 interstripe skipping peaks.
In this case, the random phase-locking degree ${\cal L}_d$, characterizing the sharpness of all the peaks, is 0.911.
Hence, the GCs make intermittent spikes which are well phase-locked to $R_{\rm GC}(t)$ at random multiples of its global period $T_G^{\rm (GC)}$.

\begin{figure}
\includegraphics[width=0.8\columnwidth]{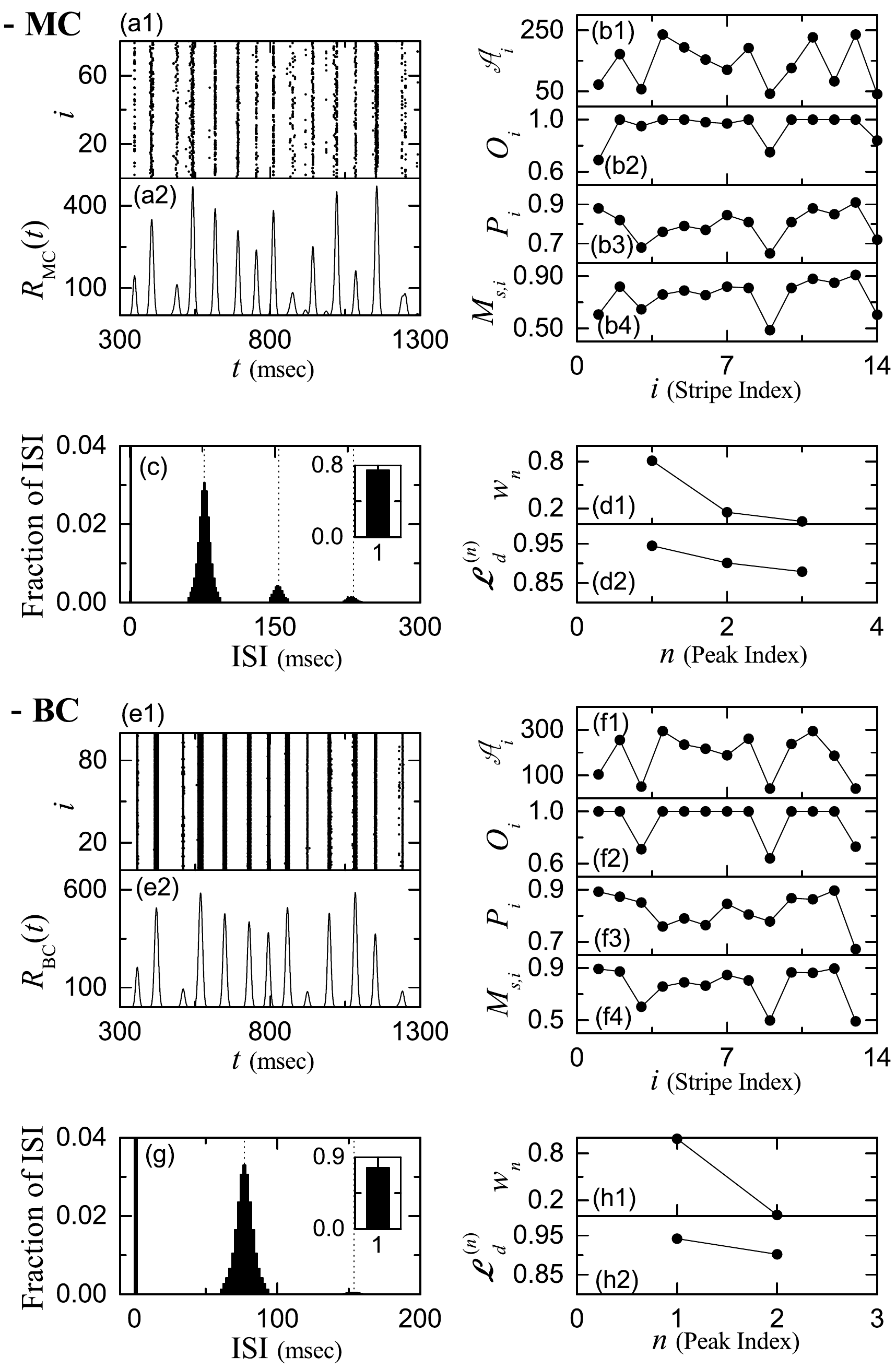}
\caption{Emergence of SSR of the MCs. (a1) Raster plot of spikes of MCs. (a2) IPSR $R_{\rm MC}(t)$ of MCs. Plots of (b1) amplitude ${\cal A}_i$ of the IPSR
$R_{\rm MC}(t)$, (b2) occupation degree $O_i$, (b3) pacing degree $P_i$, and (b4) statistical-mechanical spiking measure $M_{s,i}$ vs. $i$ (spiking stripe index). (c) Population-averaged ISI histogram. Vertical dotted lines in (c) represent the integer multiples of the global period $T_G^{\rm (MC)}$ (=76.9 msec) of $R_{\rm MC}(t)$. Plots of (d1) normalized weight $w_n$ and (d2) random phase-locking degree ${\cal L}_d^{(n)}$ for the $n$th-order peak of the ISI histogram versus $n$
(skipping peak index). Emergence of SSR of the BCs. (e1) Raster plot of spikes of BCs. (e2) IPSR $R_{\rm BC}(t)$ of BCs. Plots of (f1) amplitude ${\cal A}_i$ of the IPSR $R_{\rm BC}(t)$, (f2) occupation degree $O_i$, (f3) pacing degree $P_i$, and (f4) statistical-mechanical spiking measure $M_{s,i}$ vs. $i$ (spiking stripe index). (g) Population-averaged ISI histogram. Vertical dotted lines in (g) represent the integer multiples of the global period $T_G^{\rm (BC)}$ (=76.9 msec) of $R_{\rm BC}(t)$. Plots of (h1) normalized weight $w_n$ and (h2) random phase-locking degree ${\cal L}_d^{(n)}$ for the $n$th-order peak of the ISI histogram versus $n$
(skipping peak index).
}
\label{fig:MC-BC}
\end{figure}

\subsection{Population and Individual Behaviors in the Sparsely Synchronized Rhythms of the MCs and The BCs}
\label{subsec:MCBCR}
In our DG network, the GCs and the hilar MCs are mutually connected with the $20 \%$ random connection probabilities $p^{\rm (MC,GC)}$ ($\rm GC \rightarrow MC$) and $p^{\rm (GC,MC)}$ ($\rm MC \rightarrow GC$), which leads to formation of the GC-MC dynamical loop. Then, SSR emerges in the population of the MCs via interaction with the GCs. Also, each BC receives excitation from all the GCs in the same GC cluster, and it provides feedback inhibition to all the GCs. Thus, the GC-BC dynamical loop is formed, and SSR appears in the population of the BCs through interaction with the GCs.

Here, we investigate the population and individual firing behaviors in the SSRs of the MCs and the BCs. Unlike the case of the GCs, all the MCs ($N_{\rm MC}=80$) and all the BCs ($N_{\rm BC}=100$) are active ones (i.e., their activation degrees $D_a$ are $100 \%$) \cite{WTA}.
Their raster plots of spikes and the corresponding IPSRs [i.e., $R_{\rm MC}(t)$ and $R_{\rm BC}(t)$] are shown in Figs.~\ref{fig:MC-BC}(a1)-\ref{fig:MC-BC}(a2) and
Figs.~\ref{fig:MC-BC}(e1)-\ref{fig:MC-BC}(e2), respectively. As in the case of the GCs, SSRs with the population frequency $f_p^{(X)}~(=13$ Hz; $X=$ MC and BC) appear in the populations of the MCs and the BCs via interaction in the GC-MC-BC loop, respectively.

We note that the population frequencies of the IPSRs $R_X(t)$ ($X=$ GC, MC, and BC) are the same through mutual interaction in the GC-MC-BC loop; for convenience, sometimes we denote the population frequency just as $f_p$ without the superscript. However, phase shifts between the SSRs occur as follows. With respect to the excitatory EC input, starting at $t=300$ msec, the firings of the GCs begin at a delayed time $t = 340$ msec. The GCs provide the excitatory inputs to the MCs which then give the excitatory inputs to the BCs.
Thus, time-delay occurs for the firings of the MCs and the BCs with respect to the firings of the GCs.
This delayed firing of the MCs (BCs) may be seen clearly in the cross-correlation between the IPSR $R_{\rm MC}(t)$ [$R_{\rm BC}(t)$] and
$R_{\rm GC}(t)$;
\begin{equation}
{\cal{C}}_{X - \rm GC} (\tau) = \frac{\overline{\Delta R_{\rm GC}(t+\tau) \Delta R_{X}(t)}}{\sqrt{\overline{\Delta R_{\rm GC}^{2}(t)}} \sqrt{\overline{\Delta R_{X}^2(t)}}}; ~~~ X = {\rm MC ~ or ~ BC},
\label{eq:CC}
\end{equation}
where $\Delta R_{\rm GC}(t) = R_{\rm GC}(t)-\overline{R_{\rm GC}(t)}$, $\Delta R_{X}(t) = R_{X}(t)-\overline{R_{X}(t)}$, and the overline denotes the time average.
It is thus found that ${\cal{C}}_{\rm MC-GC}(t)$ and ${\cal{C}}_{\rm BC-GC}(t)$ have the maxima at $\tau = 10 $ and 20 msec, respectively.
Consequently, the MCs and the BCs begin to fire at delayed time $t = 350$ and 360 msec, respectively.

As in the case of the GCs, we characterize population firing behaviors in the SSRs of the MCs and the BCs.
We first employ the thermodynamic amplitude measure ${\cal M}_a$, given by the time-averaged amplitude of the macroscopic IPSRs, $R_{\rm MC}(t)$ and $R_{\rm BC}(t)$
\cite{AM}.  Figures \ref{fig:MC-BC}(b1) and \ref{fig:MC-BC}(f1) show the plots of the amplitude ${\cal A}_i$ versus $i$ (spiking stripe index) in the case of the MCs and the BCs, respectively. We follow the 389 stripes during the stimulus time $T_s$ ($= 3 \cdot 10^4$ msec), and thus the thermodynamic amplitude measures ${\cal M}_a$ (corresponding to the time-averaged amplitude $\overline { {\cal A}_i }$) for the MCs and the BCs are found to be 99.12 and 112.75, respectively, which are mush larger than ${\cal M}_a$ (=3.57) for the GCs. Hence, the synchronization degrees of the SSRs for the MCs and the BCs are much higher (about 30 times) than that for the GCs.

Next, we use the occupation degree $O_i$, the pacing degree $P_i$, and the statistical-mechanical spiking measure $M_{s,i}$ \cite{SM} for characterization
of the population firing behaviors in the SSRs of the MCs and the BCs \cite{SM}. Figures \ref{fig:MC-BC}(b2)-\ref{fig:MC-BC}(b4) and Figures \ref{fig:MC-BC}(f2)-\ref{fig:MC-BC}(f4) show the plots of $O_i$, $P_i,$ and $M_{s,i}$ in the $i$th spiking stripes for the MCs and the BCs, respectively. We follow the 389 stripes during the stimulus period $T_s$ ($= 3 \cdot 10^4$ msec), and get the average occupation $\langle O_i \rangle$, the average pacing degree $\langle P_i \rangle$, and the statistical-mechanical spiking measure $M_s$. The average occupation degrees $\langle O_i \rangle$ of the MCs and the BCs are 0.86 and 0.92, respectively, which are much larger (about 6 times) than that (= 0.14) of the GCs. However, since $\langle O_i \rangle$ of the MCs and the BCs are still less than 1, MCs and BCs also exhibit sparsely synchronized firings, but these firings are much less sparse than those of the GCs.

Also, the average pacing degrees $\langle P_i \rangle$ of the MCs and the BCs are 0.73 and 0.77, respectively, which are larger than that (= 0.45) of the GCs; the pacing between spikings for the MCs and the BCs are better than that for the GCs. Consequently, the statistical-mechanical spiking measure $M_s$ (representing the overall degree of population synchronization) of the MCs and the BCs are 0.63 and 0.71, respectively, which are much larger (at least 10 times) than that (= 0.063) of the GCs. As explained in Subsec.~\ref{subsec:GCR}, $M_s$ for the GCs becomes very small mainly due to low average occupation degree $\langle O_i \rangle$ (=0.14) (resulting from the sparse firings of the GCs).

In addition to the population behaviors, we also characterize individual spiking behaviors in the SSRs of the MCs and BCs in terms of their ISIs.
Figures \ref{fig:MC-BC}(c) and \ref{fig:MC-BC}(g) show the population-averaged ISI histograms for the MCs and the BCs, respectively; these ISI histograms
are obtained in the same way as in the GCs. Unlike the case of the GCs, the MCs and the BCs exhibit ``intrastripe bursting'' (corresponding to repeatedly firing bursts of spikes) within the stripes, in addition to the random spike skipping for the interstripe spikings; no intrastripe bursting occurs for the GCs.

Thus, the ISI histograms consist of the dominant ``intrastripe bursting peak'' [located near the ISI ($\simeq 1.11$ msec)], arising from the intrastripe burstings, as well as the interstripe skipping peaks [located at the integer multiples of the global period $T_G^{(X)}~(=76.9$ msec); $X=$ MC and BC], resulting from the random spike skipping for the interstripe spikings; the fractions of the ISIs at the intrastripe bursting peaks are 0.75 and 0.78 for the MCs and the BCs, respectively.
In this way, the structure of the ISI histograms for the MCs and the BCs is distinctly different from that for the GCs, due to the occurrence of intrastripe burstings. Consequently, for the MCs (BCs), the average ISI ($\langle {\rm ISI} \rangle$) is 23.7 (18.1) msec, and hence the population-averaged MFR $\langle f_i^{\rm (MC)} \rangle$ ($\langle f_i^{\rm (BC)} \rangle$) (=$1/\langle {\rm ISI} \rangle$) is 42.1 (55.4) Hz, which is higher than the population frequency $f_p~(=13$ Hz) of the SSRs, in contrast to the case of the GCs with $\langle f_i^{\rm (GC)} \rangle = 2$ Hz (much lower than $f_p$).

As in the case of the GCs, we also characterize the random spike skipping, leading to the interstripe skipping peaks in the ISI histograms, in terms of the random phase locking degree ${\cal L}_d$, representing how well intermittent interstripe skipping spikes make phase-locking to $R_X(t)$ at random multiples of its global period $T_G^{(X)}~(=76.9$ msec; $X=$ MC and BC). Unlike the case of the GCs (with the 13 peaks), only the 3 (2) interstripe skipping stripes appear for the MCs (BCs), due to appearance of the dominant intrastripe bursting peak. In this case, the normalized weight $w_n$ for the $n$th-order interstripe skipping peak is given by:
\begin{equation}
w_n = \frac {N_{\rm ISI}^{\rm (n,skip)}} {N_{\rm ISI}^{\rm (tot,skip)}},
\label{eq:Intrawn}
\end{equation}
where $N_{\rm ISI}^{\rm (tot,skip)}$ is the total number of interstripe skipping ISIs and $N_{\rm ISI}^{\rm (n,skip)}$ is the number of the ISIS in the
$n$th-order interstripe skipping peak.
Figures \ref{fig:MC-BC}(d1) and \ref{fig:MC-BC}(h1) show the plots of the normalized weights $w_n$ versus $n$ (peak index) for the MCs and the BCs, respectively. Unlike the GCs, the 1st-order skipping peak is dominant; $w_1=$ 0.81 and 0.99 for the MCs and the BCs, respectively; the weights of the remaining higher-order skipping peaks are very low.

We now examine the regularity of individual interstripe spikings (represented well in the sharpness of the interstripe skipping peaks) in terms of the
random phase-locking degree ${\cal L}_d$, introduced in Eq.~(\ref{eq:LD}). Figures \ref{fig:MC-BC}(d2) and \ref{fig:MC-BC}(h2)
show the plots of the random phase-locking degree ${\cal L}_d^{(n)}$ of the $n$th-order interstripe skipping peak versus $n$ (skipping peak index) for the MCs and the BCs, respectively. Then, ${\cal L}_d$, corresponding to the average of contributions of all the ISIs in the ISI histogram, is given by the weighted mean
of the random phase-locking degrees ${\cal L}_d^{(n)}$ of the $n$th-order interstripe skipping peak; ${\cal L}_d$ is 0.936 and 0.941 for the MCs and the BCs, respectively. Similar to the case of the GC with ${\cal L}_d$=0.911, the values of ${\cal L}_d$ are also very high, which implies that
the intermittent interstripe skipping spikes for the MCs and the BCs are well  phase-locked to $R_{X}(t)$ at random multiples of its global period $T_G^{\rm (X)}$
$(X=$ MC and BC).

\begin{figure}
\includegraphics[width=0.9\columnwidth]{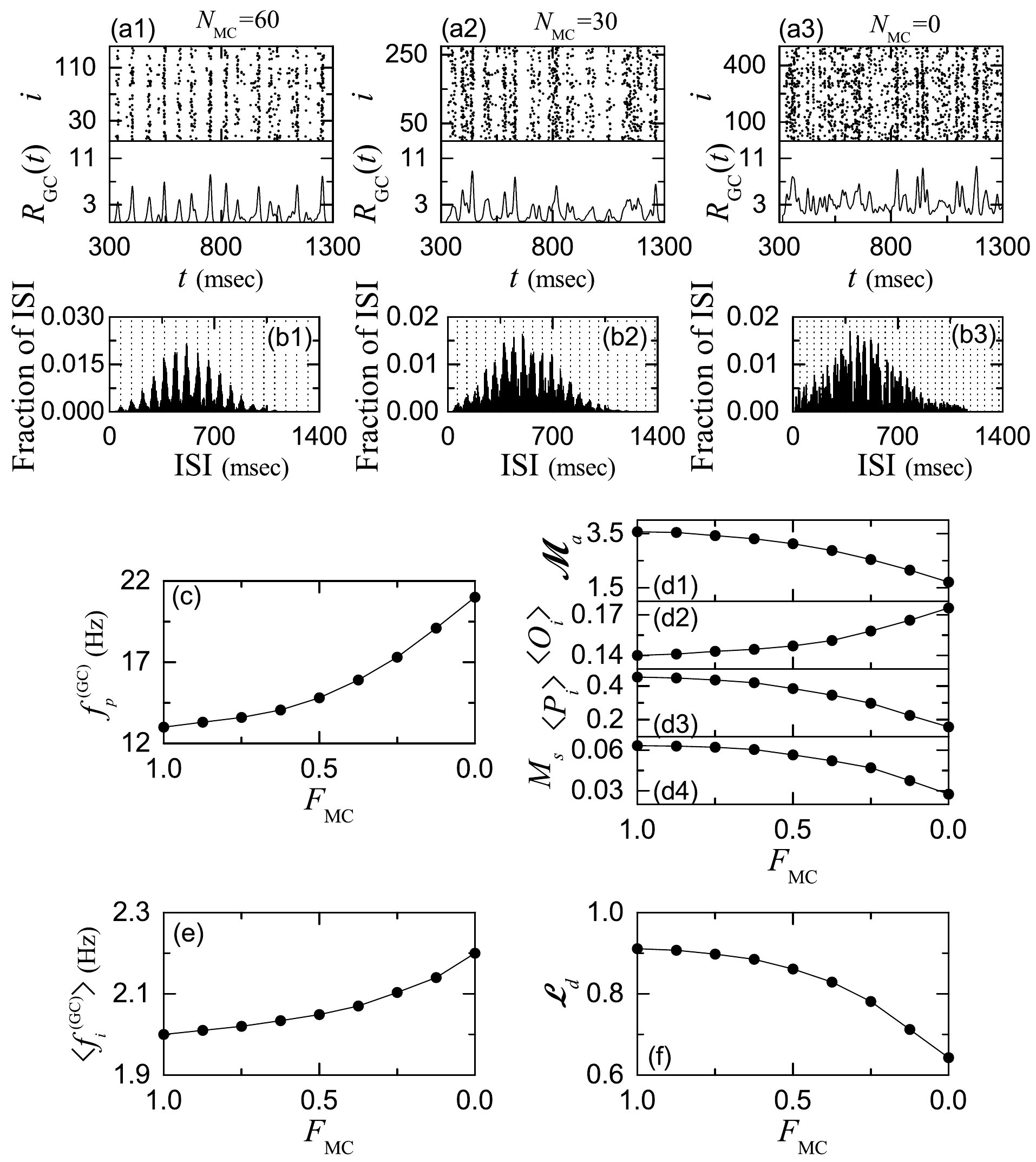}
\caption{Effect of the hilar MCs on the population and individual behaviors in the SSR of the active GCs. Raster plots of spikes and IPSRs $R_{\rm GC}(t)$ for (a1) $N_{\rm MC}=60$, (a2) $N_{\rm MC}=30$, and (a3) $N_{\rm MC}=0$. Population-averaged ISI histograms for (b1) $N_{\rm MC}=60$, (b2) $N_{\rm MC}=30$, and (b3) $N_{\rm MC}=0$. Vertical dotted lines in the ISI histograms represent the integer multiples of the global period $T_G^{\rm (GC)}$ of the SSR;
$T_G^{\rm (GC)}=$ 73.5, 62.9, 47.4 msec for $N_{\rm MC}= 60$, 30, and 0, respectively. (c) Plot of the population frequency $f_p^{\rm (GC)}$ versus
$F_{\rm MC}$ (fraction of MCs). Plots of (d1) the thermodynamic amplitude measure ${\cal M}_a$, (d2) the average occupation degree $\langle O_i \rangle$, (d3) the average pacing degree $\langle P_i \rangle$, and (d4) the statistical-mechanical spiking measure $M_s$ versus $F_{\rm MC}$. (e) Plot of the population-averaged MFRs $\langle f_i^{\rm (GC)} \rangle$  versus $F_{\rm MC}$. (f) Plot of the random phase-locking degree ${\cal L}_d$ versus $F_{\rm MC}$.
}
\label{fig:MC1}
\end{figure}

\subsection{Effect of The Hilar MCs on Population and Individual Behaviors in The Sparsely Synchronized Rhythms}
\label{subsec:MC}
The hilar MCs control the firing activity of the GC-BC loop by providing excitation to both the GCs and the BCs.
Through such control, the MCs were found to play an important role of enhancing the winner-take-all competition in each GC cluster \cite{WTA}.
However, MC loss may occur during epileptogenesis \cite{BN1,BN2}, which might be a cause of impaired pattern separation leading to memory interference.
Through ablation of the GCs, we study their effect on the firing behaviors in the SSRs of the GCs, MCs, and BCs.

We decrease $N_{\rm MC}$ (number of the MCs) from 80 (in the original whole network) to 0 (complete loss); in this case, the fraction of MCs ($F_{\rm MC}$) is given by $N_{\rm MC} / 80$. With decreasing $N_{\rm MC}$ or equivalently $F_{\rm MC}$ in the above way, we investigate change in the population and individual spiking behaviors in the SSRs of the GCs, the MCs, and the BCs, and compare them with those for $N_{\rm MC}=80$ (i.e., $F_{\rm MC}=1$) in Figs.~\ref{fig:GC} and \ref{fig:MC-BC}. It is thus found that the MCs play an essential role to enhance the synchronization degree and the random phase-locking degree in the SSRs.

We first consider the case of SSR of the GCs. Figures \ref{fig:MC1}(a1)-\ref{fig:MC1}(a3) show the raster plots of spikes and the IPSRs
$R_{\rm GC}(t)$ for $N_{\rm MC}=$ 60 ($F_{\rm MC}=0.75$), 30 ($F_{\rm MC}=0.375$), and 0 ($F_{\rm MC}=0$), respectively.
We note that sparsely synchronized spiking stripes appear successively in the rater plot of spikes and the corresponding IPSR $R_{\rm GC}(t)$ exhibits synchronous oscillations. As $N_{\rm MC}$ is decreased, the interval between the neighboring spiking stripes becomes narrower, and hence the population frequency $f_p^{\rm (GC)}$ of the SSR becomes increased.

With decreasing $N_{\rm MC}$ from 80, the firing activity of the BCs becomes weakened, which leads to decrease in the feedback inhibition to the GCs. Thus, the activation degree $D_a$ of the GCs was found to increase \cite{WTA}. Due to such increase in the firing activity of the GCs, spikes in the raster plot become more and more dense, as shown in the case of $N_{\rm MC}=$ 60, 30, and 0, which results in increase of the occupation degree $O_i$ (representing the fraction of spiking neurons in each spiking stripe). In contrast, the spiking stripes become more and more smeared, and hence the pacing degree $P_i$ (denoting the degree of phase coherence between spikes) becomes decreased.

Through competition between the occupation and the pacing degrees, the overall synchronization degree of the SSR is determined, which may be well shown in the change in the amplitude ${\cal A}_i$ of the IPSR $R_{\rm GC}(t)$; ${\cal A}_i$ in the $i$th global cycle of $R_{\rm GC}(t)$ (i.e., the $i$th spiking stripe) is given by
the difference between the maximum and the minimum of $R_{\rm GC}(t)$ divided by 2 [see Eq.~(\ref{eq:Ma})]. With decreasing $N_{\rm MC}$, the maximum $R_{\rm GC, max}^{(i)}(t)$ is found to show an increasing tendency, mainly due to the effect of the increased $O_i$; the time-averaged maximum $\overline { R_{\rm GC, max}^{(i)}(t) }=$ 7.89, 8.12, and 8.57 for $N_{\rm MC}=$ 60, 30, and 0, respectively. However, the minimum $R_{\rm GC, min}^{(i)}(t)$ exhibits more increasing tendency because of the effects of the increased $O_i$ and the decreased $P_i$; the time-averaged minimum $\overline{R^{(i)}_{\rm GC, min}(t)}=$ 1.04, 2.38, and 5.15 for $N_{\rm MC}=$ 60, 30, and 0, respectively. Consequently, with decreasing $N_{\rm MC}$ the thermodynamic amplitude measure ${\cal M}_a$ (representing the time-averaged amplitude) becomes decreased (i.e., the overall synchronization degree decreases).

By decreasing $F_{\rm MC}$ from 1 to 0, we make more quantitative characterization of the population firing behavior for various values of $F_{\rm MC}$. Figure \ref{fig:MC1}(c) and Figs.~\ref{fig:MC1}(d1)-\ref{fig:MC1}(d4) show the plot of the population frequency $f_p^{\rm (GC)}$ versus $F_{\rm MC}$ and the plots of the amplitude measure ${\cal M}_a$, the average occupation degree $\langle O_i \rangle$,  the average pacing degree $\langle P_i \rangle$, and the
statistical-mechanical spiking measure $M_s$ versus $F_{\rm MC}$, respectively; all these quantities are obtained by following all the spiking stripes appearing during the stimulus period $T_s$ (= $3 \cdot 10^4$ msec). As a result of the increased firing activity of the GCs, the population frequency $f_p^{\rm (GC)}$ is found to increase from 13.0 to 21.1 Hz [see Fig.~\ref{fig:MC1}(c)]. The frequency range of the SSR corresponds to the beta rhythm.

\begin{figure}
\includegraphics[width=0.9\columnwidth]{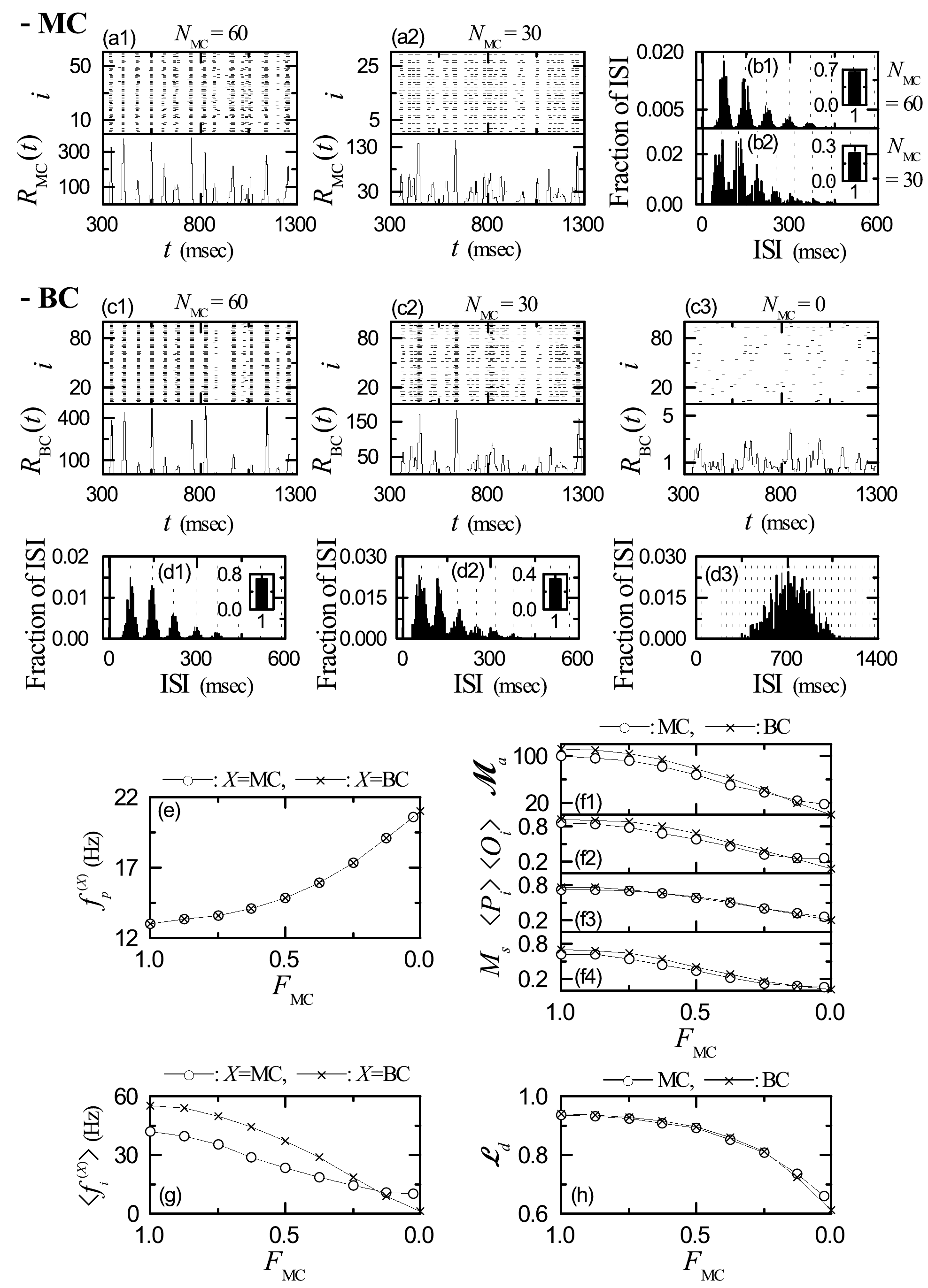}
\caption{Effect of the hilar MCs on the population and individual behaviors in the SSRs of the MCs and the BCs. Raster plots of spikes of the MCs and IPSR $R_{\rm MC}(t)$ for (a1) $N_{\rm MC}=60$ and (a2) $N_{\rm MC}=30$. Population-averaged ISI histograms for the MCs for (b1) $N_{\rm MC}=60$ and (b2) $N_{\rm MC}=30$. Vertical dotted lines in the ISI histograms represent the integer multiples of the global period $T_G^{\rm (MC)}$ of $R_{\rm MC}(t)$; $T_G^{\rm (MC)}=$ 73.5 and 62.9 msec for $N_{\rm MC}=$ 60 and 30, respectively.
Raster plots of spikes of the BCs and IPSR $R_{\rm BC}(t)$ for (c1) $N_{\rm MC}=60,$ (c2) $N_{\rm MC}=30$, and (c3) $N_{\rm MC}=0$. Population-averaged ISI histograms for the BCs for (d1) $N_{\rm MC}=60,$ (d2) $N_{\rm MC}=30,$ and (d3) $N_{\rm MC}=0$. Vertical dotted lines in the ISI histograms represent the integer multiples of the global period $T_G^{\rm (BC)}$ of $R_{\rm BC}(t)$; $T_G^{\rm (BC)}=$ 73.5, 62.9, and 47.4  msec for $N_{\rm MC}=$ 60, 30, and 0, respectively.
(e) Plot of the population frequency $f_p^{(X)}$ versus $F_{\rm MC}$ (fraction of MCs); $X=$ MC (open circle) and BC (cross). Plots of (f1) the thermodynamic amplitude measure ${\cal M}_a$, (f2) the average occupation degree $\langle O_i \rangle$, (f3) the average pacing degree $\langle P_i \rangle$, and (f4) the statistical-mechanical spiking measure $M_s$ versus $F_{\rm MC}$; MC (open circle) and BC (cross).
(g) Plot of the population-averaged MFRs $\langle f_i^{(X)} \rangle$  versus $F_{\rm MC}$; $X=$ MC (open circle) and BC (cross).
(h) Plot of the random phase-locking degree ${\cal L}_d$ versus $F_{\rm MC}$; MC (open circle) and BC (cross).
}
\label{fig:MC2}
\end{figure}

The synchronization degree of the SSR with the beta-range $f_p^{\rm (GC)}$ is characterized in terms of the thermodynamic amplitude measure ${\cal M}_a$ and the statistical-mechanical spiking measure $M_s$. As $F_{\rm MC}$ is decreased from 1 to 0, ${\cal M}_a,$ [representing the time-averaged amplitude of $R_{\rm GC}(t)$], is found to decrease from 3.57 to 1.71, as shown in Fig.~\ref{fig:MC1}(d1). Hence, the overall synchronization degree of the SSR becomes decreased. Due to increase in the firing activity of the GCs with decreasing $F_{\rm MC}$, the average occupation degree $\langle O_i \rangle$ of the spikes becomes increased from 0.14 to 0.18; more dense spikes appear in the raster plot. In contrast, as $F_{\rm MC}$ is decreased, the average pacing degree $\langle P_i \rangle$ between the dense spikes is found to decrease from 0.45 to 0.16. Then, the overall synchronization degree of the SSR is determined through competition between the (increasing) occupation and the (decreasing) pacing degrees. In this case, the pacing between the dense spikes becomes much worse, and hence the statistical-mechanical spiking measure $M_s$, given by the product of the occupation and the pacing degrees, is found to decrease from 0.063 to 0.028, which is in consistent with the decrease in ${\cal M}_a$.

Next, we consider the individual firing behavior in the SSR of the GCs. Figures \ref{fig:MC1}(b1)-\ref{fig:MC1}(b3) show the ISI histograms for
$N_{\rm MC}=$ 60, 30, and 0, respectively. The GCs exhibit intermittent interstripe spikings, locked to $R_{\rm GC}(t)$ at random multiples of the global
period $T_G^{\rm (GC)}$ of $R_{\rm GC}(t)$. Due to random spike skipping, the ISI histograms consist of the interstripe skipping peaks.
The mean ISIs ($\langle {\rm ISI} \rangle$) are 495, 483, and 455 msec, respectively, in the case of $N_{\rm MC}=$ 60, 30, and 0. Hence, the corresponding population-averaged MFRs $\langle f_i^{(GC)} \rangle$ (= $1~ /~ \langle {\rm ISI} \rangle$) are 2.02, 2.07, and 2.20 Hz, respectively, due to increased activity of the GCs. Moreover, we also note that, as $N_{\rm MC}$ is decreased, the interstripe skipping peaks become more and more smeared, which results in decrease in the random phase-locking degree ${\cal L}_d$ [representing the degree of random phase-locking to $R_{\rm GC}(t)$].

As in the case of the population behavior, with decreasing $F_{\rm MC}$ from 1 to 0, we make more quantitative characterization of the individual firing behavior for various values of $F_{\rm MC}$. Figures \ref{fig:MC1}(e) and \ref{fig:MC1}(f) show the plots of the population-averaged MFR $\langle f_i^{(GC)} \rangle$ and the random phase-locking degree ${\cal L}_d$, respectively. As $F_{\rm MC}$ is decreased from 1 to 0, $\langle f_i^{\rm (GC)} \rangle$ (given by the reciprocal of the mean ISI) is found to increase from 2.0 to 2.2 Hz, because of the increased firing activity of the GCs. We note that $\langle f_i^{(GC)} \rangle$ is much less than the population frequency $f_p^{\rm (GC)}$, due to random spike skipping. Also, the random phase-locking degree ${\cal L}_d$ (characterizing the degree of random spike skipping) is found to exhibit decreasing tendency from 0.91 to 0.64, due to smearing of the interstripe skipping peaks, as in the decrease in the population synchronization degrees, ${\cal M}_a$ and $M_s$. 

From now on, with decreasing $N_{\rm MC}$, we study the population and individual firing behaviors in the SSRs of the MCs and the BCs.
We first consider the cases of $N_{\rm MC}=$ 60, 30, and 0; $N_{\rm MC}=0$ may apply to only the case of the BCs.
Figures \ref{fig:MC2}(a1)-\ref{fig:MC2}(a2) and Figs.~\ref{fig:MC2}(c1)-\ref{fig:MC2}(c3) show the raster plots of spikes and the IPSRs
$R_{X}(t)$ ($X=$ MC and BC) for the MCs and the BCs, respectively.
As in the case of the GCs, sparsely synchronized spiking stripes appear successively in the rater plots of spikes and the corresponding IPSRs
exhibit synchronous oscillations. As $N_{\rm MC}$ is decreased, the interval between the neighboring spiking stripes becomes narrower, and hence the population frequency $f_p^{\rm (X)}$ of the SSR becomes increased.

With decreasing $N_{\rm MC}$ from 80, the firing activities of both the MCs and the BCs become weakened.
Hence, unlike the case of the GCs, spikes in the raster plot become more and more sparse, which leads to decrease in the occupation degree $O_i$ (denoting the fraction of spiking neurons in each spiking stripe). Moreover, the spiking stripes become more and more smeared, and hence the pacing degree $P_i$
(representing the degree of phase coherence between spikes) also becomes decreased, as in the case of the GCs.

The overall synchronization degree of the SSRs may be characterized in terms of the thermodynamic amplitude measure ${\cal M}_a$, given by the time-averaged amplitude $\overline {{\cal A}_i}$ of the IPSR $R_X(t)$; ${\cal A}_i$ is given by the difference between the maximum and the minimum of $R_X(t)$ divided by 2.
Unlike the case of the GCs, with decreasing $N_{\rm MC}$, the maximum $R_{\rm GC, max}^{(i)}(t)$ is found to show a decreasing tendency, mainly due to the effect of the decreased $O_i$. Due to decrease in $P_i$, $R_{\rm GC, max}^{(i)}(t)$ becomes more decreased. The minimum $R_{\rm GC, min}^{(i)}(t)$ is also found to decrease due to decrease in $O_i$. However, because of decrease in $P_i$, $R_{\rm GC, min}^{(i)}(t)$ becomes less decreased. As a result, as $N_{\rm MC}$ is decreased, the  
amplitude measure ${\cal M}_a$ becomes decreased, as in the case of the GCs. Thus, the synchronization degrees of all the SSRs for the GCs, the MCs, and the BCs decrease with decreasing $N_{\rm MC}$.

Moreover, by decreasing $F_{\rm MC}$ from 1 to 0, we make more quantitative characterization of the population firing behavior for various values of $F_{\rm MC}$;
for the MCs, instead of $N_{\rm MC}=0$, we consider the case of $N_{\rm MC}=2$ (i.e., $F_{\rm MC}= 0.025$) which corresponds to the simplest coupled case. All relevant quantities are obtained by following all the spiking stripes appearing during the stimulus period $T_s$ (= $3 \cdot 10^4$ msec). Figure \ref{fig:MC2}(e) shows the plot of the population frequency $f_p^{(X)}$ [$X=$ MC (open circle) and BC (cross)] versus $F_{\rm MC}$. Through interaction with the GCs in the GC-MC and the GC-BC loops, the population frequencies $f_p^{(X)}$ of the SSRs for the MCs and the BCs  are found to increase from 13 to 21.1 Hz in the same way as that for the GCs in Fig.~\ref{fig:MC1}(c). Thus, the GCs, the MCs, and the BCs exhibit SSRs with the same beta-range population frequency $f_p$ which increases with decreasing
$F_{\rm MC}$.

The synchronization degree of the SSR for the MCs and the BCs may be characterized in terms of the thermodynamic amplitude measure ${\cal M}_a$ and the statistical-mechanical spiking measure $M_s$. Figures~\ref{fig:MC2}(f1)-\ref{fig:MC2}(f4) show the plots of the amplitude measure ${\cal M}_a$, the average occupation degree $\langle O_i \rangle$, the average pacing degree $\langle P_i \rangle$, and the statistical-mechanical spiking measure $M_s$ versus $F_{\rm MC}$, respectively; MC (open circle) and BC (cross). As $F_{\rm MC}$  is decreased from 1 to 0, the thermodynamic amplitude measure ${\cal M}_a$ [denoting the time-averaged amplitude of $R_X(t)$ ($X=$ MC and BC)] for the MCs (BCs) is found to decrease rapidly from 99.1 (112.8) to 18.6 (0.98), as in the case of the GCs.

Unlike the case of the GCs, due to decreased firing activity of the MCs (BCs) with decreasing $F_{\rm MC}$ from 1 to 0, the average occupation degree $\langle O_i \rangle$ of the spikes becomes decreased from 0.86 (0.92) to 0.25 (0.09); more sparse spikes appear in the raster plot.
Similarly, with decreasing $F_{\rm MC}$, the average pacing degree $\langle P_i \rangle$ between the sparse spikes is also found to decrease from 0.73 (0.77) to 0.26 (0.21) for the MCs (BCs). Then, the statistical-mechanical spiking measure $M_s$ for the MCs (BCs), given by the product of the occupation and the pacing degrees, is found to decrease from 0.63 (0.71) to 0.065 (0.019), which is in consistent with the decrease in ${\cal M}_a$. In this way, as $F_{\rm MC}$ is decreased, the synchronization degrees of the SSRs for the GCs, the MCs, and the BCs become decreased together.

In addition to the population behaviors, we also study the individual firing behaviors in the SSRs of the MCs and the BCs.
Figures \ref{fig:MC2}(b1)-\ref{fig:MC2}(b2) and Figs.~\ref{fig:MC2}(d1)-\ref{fig:MC2}(d3) show the ISI histograms for
the MCs and the BCs, respectively. Unlike the GCs, both the MCs and the BCs exhibit intrastripe burstings as well as interstripe intermittent spikings, locked to
the IPSR $R_{X}(t)$ ($X=$ MC and BC) at random multiples of the global period $T_G^{(X)}$ of $R_{X}(t)$.
As a result, the ISI histograms for the MCs and the BCs consist of both the intrastripe bursting peak and the interstripe skipping peaks.

We note that, with decreasing $N_{\rm MC}$, the intrastripe bursting activity of the MCs and the BCs becomes weakened, and hence the height of the intrastripe bursting peak becomes decreased, which results in development of the interstripe skipping peaks; e.g., more and more higher-order skipping peaks appear, in comparison to the case of $N_{\rm MC}=80$ in Fig.~\ref{fig:MC-BC}. Particularly, in the case of the BCs, when passing $N_{\rm MC}=9$ (i.e., $F_{\rm MC}=0.1125$), intrastripe bursting peak is found to disappear; for the MCs intrastripe bursting peak persists until $N_{\rm MC}=1$. Thus, only the interstripe skipping peaks appear for $N_{\rm MC}=0$ in Fig.~\ref{fig:MC2}(d3). In this way, with decreasing $N_{\rm MC}$ the interstripe skipping peaks become more and more developed; particularly, more development occurs for the BCs than the MCs. Hence, the mean ISIs ($\langle {\rm ISI} \rangle$) become increased for the MCs and the BCs, which results in decrease in the population-averaged MFRs $\langle f_i^{(X)} \rangle$ ($X=$ MC and BC).

Thus, as $F_{\rm MC}$ is decreased from 1 to 0.025 (0) for the MCs (BCs), $\langle f_i^{(X)} \rangle$ ($X=$ MC and BC) is found to decrease from 42.1 (55.4) to 10.5 (1.3) Hz [see Fig.~\ref{fig:MC2}(g)], in contrast to the increase in $\langle f_i^{\rm (GC)} \rangle$ for the GCs in Fig.~\ref{fig:MC1}(e).
Unlike the case of the GCs, for large $F_{\rm MC}$ with strong intrastripe burstings, the population-averaged MFRs $\langle f_i^{(X)} \rangle$ are higher than the population frequency $f_p$. However, as $F_{\rm MC}$ is decreased, the intrastripe bursting activity becomes decreased, and then the interstripe skipping activity
becomes intensified. Then, for small $F_{\rm MC}$ with strong interstripe skipping activity (i.e., intrastripe burstings is very weak),
$\langle f_i^{(X)} \rangle$ becomes less than $f_p$, as in the case of the GCs (without intrastripe burstings).

We also note that, as $F_{\rm MC}$ is decreased, the interstripe skipping peaks become more and more smeared, which leads to decrease in the random phase-locking degree ${\cal L}_d$ [denoting the degree of random phase-locking to $R_X(t)$]. With decreasing $F_{\rm MC}$ from 1 to 0.025 (0) for the MCs (BCs),
${\cal L}_d$ is found to decrease from 0.936 (0.941) to 0.662 (0.613), as shown in Fig.~\ref{fig:MC2}(h), as in the case of the GCs.
Thus, as $F_{\rm MC}$ is decreased, the random phase-locking degrees for the GCs, the MCs, and the BCs become decreased together, as in the synchronization degrees
of the SSRs.

Finally, we compare the firing behaviors between the MCs and the BCs.
For large $F_{\rm MC}$, the BCs exhibit firing activity with higher MFR $\langle f_i^{\rm (BC)} \rangle$ than the MCs, due to stronger intraburst bursting
activity. However, as $F_{\rm MC}$ is sufficiently  decreased, the bursting activity of the BCs becomes very weak due to weak excitation from the MCs.
Then, $\langle f_i^{\rm (BC)} \rangle$ begins to decrease so rapidly and it becomes lower than $\langle f_i^{\rm (MC)} \rangle$ [see Fig.~\ref{fig:MC2}(g)].
Similarly, for large $F_{\rm MC}$ with strong intrastripe bursting activity, the random phase-locking degree ${\cal L}_d$ for the BCs is also a little larger than that for the MCs, while for sufficiently small $F_{\rm MC}$ with so weak intrastripe bursting activity, ${\cal L}_d$ for the BCs decreases rapidly and it becomes less than that for the MCs, as shown in Fig.~\ref{fig:MC2}(h). The population firing behaviors for the MCs and the BCs are also similarly as follows. For large $F_{\rm MC}$ with strong intrastripe bursting burst activity, ${\cal M}_a$, $\langle O_i \rangle$, $\langle P_i \rangle$, and $M_s$ for the BCs are larger than those for the MCs, while for sufficiently small $F_{\rm MC}$ (with very weak intrastripe bursting activity), those for the MCs become larger than those for the BCs
[see Figs.~\ref{fig:MC2}(f1)-\ref{fig:MC2}(f4)].

\begin{figure}
\includegraphics[width=0.9\columnwidth]{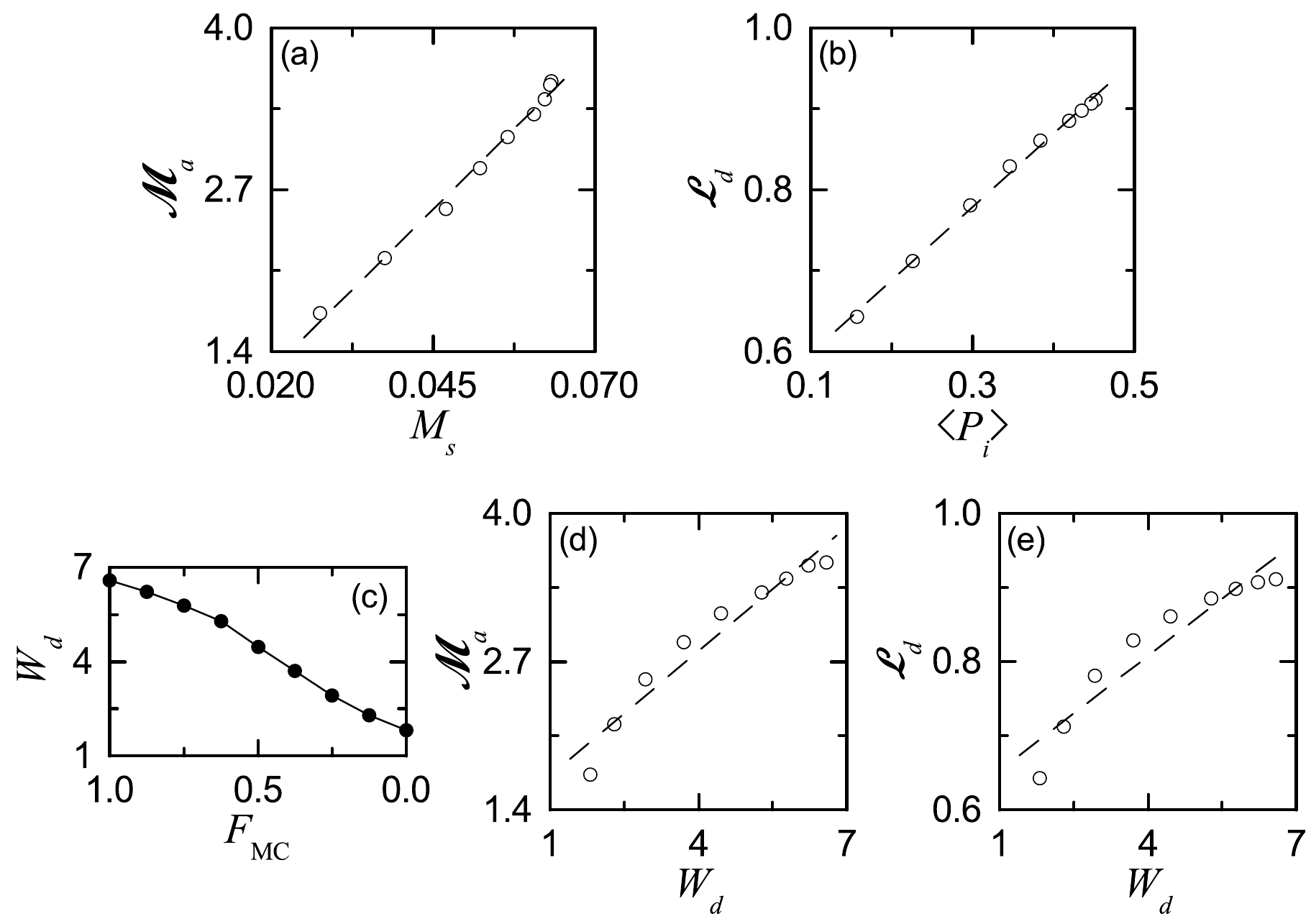}
\caption{Quantitative relationship between the SSR and the winner-take-all competition.
(a) Plot of the thermodynamic amplitude measure ${\cal M}_a$ versus the statistical-mechanical spiking measure $M_s$.
(b) Plot of the random phase-locking degree ${\cal L}_d$ versus the average pacing degree $\langle P_i \rangle$.
(c) Plot of the winner-take-all competition degree $W_d$ versus $F_{\rm MC}$ (fraction of the MCs).
(d) Plot of ${\cal M}_a$ versus $W_d$.
(e) Plot of ${\cal L}_d$ versus $W_d$.
Fitted dashed lines are given in (a)-(e). 
}
\label{fig:MC3}
\end{figure}

\subsection{Quantitative Relationship between The Sparsely Synchronized Rhythm and The Winner-Take-All Competition}
\label{subsec:Corr}
The main encoding GCs was found to exhibit sparse activation via winner-take-all competition in each GC cluster; only strongly active GCs survive
under the feedback inhibition of the BC \cite{WTA}. Such sparsity was thought to improve the pattern separation in the DG.
In Subsec.~\ref{subsec:GCR}, SSR is found to appear in the population of the GCs, along with occurrence of the winner-take-all competition.
We investigate the quantitative association between the SSR and the winner-take-all competition.

We first consider the association between the measures characterizing the SSR of the GCs.
Figure \ref{fig:MC3}(a) shows the plot of the thermodynamic amplitude measure ${\cal M}_a$ versus the statistical-mechanical spiking measure
$M_s$; plots of ${\cal M}_a$ and $M_s$ versus $F_{\rm MC}$ are shown in Figs.~\ref{fig:MC1}(d1) and \ref{fig:MC1}(d4), respectively.
The thermodynamic and statistical-mechanical synchronization degrees, ${\cal M}_a$ and $M_s$, (characterizing the population firing behavior in the SSR) are strongly correlated with the Pearson's correlation coefficient $r=0.9957$ \cite{Pearson}.

Individual firing behaviors of the GCs are characterized in terms of the ISIs. Due to the random spike skipping, the ISI histogram consists of the
interstripe skipping peaks. The random phase-locking degree ${\cal L}_d$ is used to  characterize the degree of random spike skipping (i.e., degree of sharpness of the interstripe skipping peaks in the ISI histogram). Figure \ref{fig:MC3}(b) shows the plot of ${\cal L}_d$ versus the average pacing degree $\langle P_i \rangle$;
plots of ${\cal L}_d$ and $\langle P_i \rangle$ versus $F_{\rm MC}$ are shown in Figs.~\ref{fig:MC1}(f) and \ref{fig:MC1}(d3), respectively.
${\cal L}_d$ is found to be strongly correlated with $\langle P_i \rangle$ (characterizing the smearing degree of the spiking stripes in the raster plot of spikes)
with the Pearson's correlation coefficient $r=0.9983$.

We now consider the winner-take-all competition occurring in each GC cluster via competition between the firing activity of the GCs and the feedback inhibition of the BC; for details refer to \cite{WTA}. The firing activity of the GCs is determined through competition between the external excitatory (E) to inhibitory (I) inputs to the GCs; two types of E inputs from the EC cells and the MCs and one kind of I input from the HIPP cells.
The E-I conductance ratio ${ {\cal R}_{\rm E-I}^{\rm (con)} }^*$ (given by the time-average of the external E to I conductances) was found to represent well the degree of the external E-I input competition. GCs with larger ${ {\cal R}_{\rm E-I}^{\rm (con)} }^*$ than a threshold survived in response to the feedback
of the BC (i.e., they became winners). It was thus shown that GCs become active winners when their ${ {\cal R}_{\rm E-I}^{\rm (con)} }^*$ lies within the winner
threshold percentage $W_{th}\%$ of the maximum  ${ {\cal R}_{\rm E-I}^{\rm (con)} }^*$(max) of the GC with the strongest activity; see Eq. (23) for $W_{th}\%$
in \cite{WTA}. As $F_{\rm MC}$ is decreased from 1 to 0, $W_{th}\%$ was found to increase from 15.1 $\%$ to 55 $\%$ (see Fig. 10(f2) in \cite{WTA}).
Due to the increased $W_{th}\%$, more active GCs appear with decreasing $F_{\rm MC}$, and hence the winner-take-all competition becomes weaker.

Here, we introduce the winner-take-all competition degree $W_d$ which is reciprocally related to the winner threshold percentage $W_{th}\%$:
\begin{equation}
W_d = {\frac {100} {W_{th}\%}}.
\label{eq:Wd}
\end{equation}
The smaller $W_{th}\%$ is, the larger $W_d$ is.
Figure \ref{fig:MC3}(c) shows the plot of $W_d$ versus $F_{\rm MC}$.
With decreasing $F_{\rm MC}$, $W_d$ is decreased, and hence the winner-take-all competition becomes weaker.

Figures \ref{fig:MC3}(d) and \ref{fig:MC3}(e) show plots of ${\cal M}_a$ and ${\cal L}_d$ versus $W_d$, respectively.
Population (${\cal M}_a$) and individual (${\cal L}_d$) firing behaviors in the SSR are found to be positively
correlated with the winner-take-all competition ($W_d$) with the Pearson's correlation coefficients $r=0.9709$ and 0.9599, respectively.
Hence, the synchronization and the random phase-locking degrees in the SSR of the GCs become higher when the winner-take-all competition is stronger. 

\section{Summary and Discussion}
\label{sec:SUM}
We investigated population and individual behaviors in the SSRs in a spiking neural network of the hippocampal DG.
Through interaction of excitation of the GCs with inhibition of the BCs, SSRs have been found to appear in each population of the GCs and the BCs,
along with occurrence of the winner-take-all competition in each GC cluster, leading to sparse activation of the GCs.
Such sparsity has been known to be directly associated with pattern separation, facilitating pattern storage and retrieval in the area CA3.

In each case of the GCs and the BCs, sparsely synchronized stripes have been found to appear successively in the raster plots of spikes,
and the corresponding IPSR $R_X(t)$ ($X=$ GC and BC) exhibited oscillatory behavior with the population frequency $f_p~ (= 13$ Hz).
Such SSR has also been found to appear in the population of the hilar MCs (controlling the firing activity of the GC-BC loop) via interaction
with the GCs in the GC-MC loop. Thus, SSRs of the GCs, the MCs, and the BCs emerged in the whole DG network.
Similar sparsely-synchronized population rhythms were observed in the hippocampus, the neocortex, the cerebellum, and the olfactory system \cite{FSS1,FSS2,FSS3,FSS4}.

We have made intensive characterization of the population behaviors in the SSRs of the GCs, the MCs, and the BCs by employing the following diverse synchronization measures introduced in our prior works. As a thermodynamic synchronization degree, we used the amplitude measure ${\cal M}_a$, given by the time-averaged amplitude of the macroscopic IPSR $R_X(t)$ ($X=$ GC, MC, and BC) \cite{AM}, and characterized the overall synchronization degree of the SSRs. The SSR of the GCs was found to have the lowest amplitude measure (${\cal M}_a$ = 3.57, 99.12, and 112.75 for the GCs, MCs, and BCs, respectively). Next, we also made characterization of the population behaviors in terms of the statistical-mechanical spiking measure $M_s$ (based on the microscopic spikes in the raster plot), given by the product of the occupation degree $\langle O_i \rangle$ and the pacing degree $\langle P_i \rangle$ \cite{SM}. Among the 3 SSRs, the SSR of the GCs was the most sparse, because its occupation degree $\langle O_i \rangle$ (= 0.14) was so much less than those in the SSRs of the MCs and the BCs; $\langle O_i \rangle=$ 0.86 and 0.92 for the MCs and the BCs, respectively. Also, its pacing degree $\langle P_i \rangle$ between the spikes in the raster plot was lower than those for the MCs and the BCs;
$\langle P_i \rangle$ = 0.45, 0.73, and 0.77, respectively. Consequently, the statistical-mechanical spiking measure $M_s$ of the SSR for the main encoding GCs became the lowest ($M_s$ = 0.063, 0.63, and 0.71 for the GCs, MCs and the BCs, respectively), mainly due to sparse firing of the GCs (resulting from the winner-take-all competition).

In addition to the population behaviors, we have also investigated individual firing activities in the SSRs of the GCs, MCs, and GCs.
In the case of GCs, active GCs exhibited intermittent spikings, phase-locked to the IPSR $R_{\rm GC}(t)$ at random multiples of its global period $T_G~(=76.9$ msec). Due to the random spike skipping, the ISI histogram has been found to consist of distinct multiple peaks (called the interstripe skipping peaks) at the integer multiples of $T_G$, similar to the cases of previously-found ``standard'' sparse synchronization \cite{W_Review,Sparse1,Sparse2,Sparse3,FSS}. However, unlike the standard sparse synchronization where the 1st-order peak was the highest one, the middle 6th- and 7th-order peaks were the highest ones. In this multi-peaked ISI histogram, the mean ISI ($\langle {\rm ISI} \rangle$) was 500.01 msec. Then, the population-averaged MFR of the GCs $\langle f_i^{(\rm GC)} \rangle$ (=$1/ \langle {\rm ISI} \rangle$) was 2.0 Hz, which was much less than the population frequency $f_p$, mainly due to random spike skipping.

Unlike the case of GCs, MCs and BCs have been found to exhibit bursting-like multi-spikings within the stripes. Consequently, the ISI histograms for the MCs and the BCs have been found to have the intrastripe bursting peak, in addition to the interstripe skipping multi-peaks, in contrast to the standard sparse synchronization with only the interstripe skipping multi-peaks \cite{W_Review,Sparse1,Sparse2,Sparse3,FSS}. Due to the dominance of the intrastripe bursting peak,
the mean ISI ($\langle {\rm ISI} \rangle$) became shorter; $\langle {\rm ISI} \rangle=$ 23.7 and 18.1 msec for the MCs and the BCs, respectively.
Then, the population-averaged MFR for the MCs and the BCs were $\langle f_i^{(\rm GC)} \rangle$ (=$1/ \langle {\rm ISI} \rangle$) were 42.1 and 55.4 Hz, respectively, which were higher than the population frequency $f_p$ (= 13 Hz), due to the intrastripe burstings, which was in contrast to the case of the GCs
where $\langle f_i^{(\rm GC)} \rangle$ is less then $f_p$.

We also introduced a new random phase-locking degree ${\cal L}_d$ and characterized the ``sharpness'' of the interstripe skipping peaks representing how well the intermittent spikes make phase-locking to $R_{\rm GC}(t)$ at random multiples of its global period $T_G$. The random phase-locking degree ${\cal L}_d$, characterizing the degree of random spike skipping for the GCs, was a little lower than those of the MCs and the BCs; ${\cal L}_d=$ 0.911, 0.936, and 0.941 for the GCs, the MCs, and the BCs, respectively. The order in magnitude of ${\cal L}_d$ was the same as that in the synchronization degrees, ${\cal M}_a$ and $M_s$, for the SSRs.

MC loss may occur during epileptogenesis. With decreasing $F_{\rm MC}$ (fraction of the MCs) from 1 to 0, we investigated the effect of the MCs on the population and individual firing behaviors in the SSRs of the GCs, MCs, BCs. As $F_{\rm MC}$ was decreased, the interval between the spiking stripes in the raster plot
became narrowed, and the spiking stripes became more and more smeared. Hence, the population frequency $f_p$ of the SSRs showed an increasing tendency and their synchronization degrees became decreased. In the ISI histogram for the GCs, the mean ISI ($\langle {\rm ISI} \rangle$) became shorter, mainly due to weakened inhibition from the BCs, and hence the population averaged MFR $\langle f_i^{(\rm GC)} \rangle$ increased. Moreover, the interstripe skipping peaks
became more and more smeared, leading to decrease in the random phase-locking degree ${\cal L}_d$.

In the case of the MCs and the BCs, with decreasing $F_{\rm MC}$, the heights of the intrastripe bursting peaks in their ISI histograms became decreased mainly due to decrease in the firing activity of the MCs, which resulted in intensifying the interstripe skipping peaks (i.e. more and more higher-order interstripe
skipping peaks appeared). Consequently, the mean ISI ($\langle {\rm ISI} \rangle$) became longer, which led to decrease in the population-averaged MFR
$\langle f_i^{(X)} \rangle$ ($X=$ MC and BC), in contrast to the increase in $\langle f_i^{(\rm GC)} \rangle$ for the GCs. Similar to the case of the GCs, the random phase-locking degree ${\cal L}_d$ decreased because the interstripe skipping peaks became more and more smeared.

We note that the SSR of the GCs appeared along with occurrence of the winner-take-all competition in the GC clusters.
Hence, we became concerned about the quantitative correlation between the population and individual behaviors in the SSR and the winner-take-all competition for the GCs. It was thus found that both the synchronization degrees, ${\cal M}_a$ and $M_s$, and the random phase-locking degree ${\cal L}_d$ were positively correlated with the winner-take-all competition degree $W_d$.

Finally, we discuss limitations of our present work and future works.
In the present work, the population and individual behaviors in the SSRs were found to be positively correlated with the winner-take-all competition. However, this kind of correlation does not imply causal relationship. Hence, in future work, it would be interesting to make intensive investigation on their dynamical causation. Also, in the present work, we studied only the case of ablating the MCs for investigation of their role.
In future, it would also be interesting to study the population and the individual behaviors in the SSRs by varying the synaptic strength $K_R^{\rm (BC,MC)}$
$(R~=~\rm{NMDA~and~AMPA})$ of the synapse between the MC and the BC. The effect of decrease in $K_R^{\rm (BC,MC)}$ would be similar to that of
decreasing $F_{\rm MC}$, because the synaptic inputs into the BC and the GCs are decreased in both cases.

Moreover, in the present work, we took into consideration the disynaptic inhibitory effect of the MCs on the GCs (i.e., disynaptic inhibition to the GCs mediated by the BC). However, in our present DG network, we did not consider the synaptic connection from the HIPP cells to the BCs, and hence we could not
study the disynaptic effect of the HIPP cells on the GCs (i.e., HIPP $\rightarrow$ BC $\rightarrow$ GC).
The HIPP cells are known to disinhibit the BC \cite{BN1,BN2}, which results in decrease in the inhibitory effect of the BC on the GCs. Then, the activity of the GCs may increase. In this way, the disynaptic effect of the HIPP cells on the GCs, mediated by the BC, which tends to increase the activity of the GCs, is in contrast to the disynaptic inhibition of the MCs to the GCs (decreasing the firing activity of the GCs). Hence, in future work, it would be meaningful to investigate the disynaptic effect of the HIPP cells on the GC in a modified DG network (including the synaptic connections from the HIPP cells to the BCs).

\section*{Acknowledgments}
This research was supported by the Basic Science Research Program through the National Research Foundation of Korea (NRF) funded by the Ministry of Education (Grant No. 20162007688).

\end{document}